\documentstyle[12pt]{article}
\newcommand{\vf}[1]{\mbox{\boldmath $#1$}}

\begin{document}

\title{Nine lectures on quark structure of light hadrons}
\author{B.V.Martemyanov\\
\it{Institute of Theoretical and Experimental Physics}\\
\it{117218, B.Cheremushkinskaya 25, Moscow, Russia}}
\maketitle
\begin{center}
{\bf Abstract}
\end{center}

The following lectures concern with the quark structure of
hadrons made up from light quarks. The symmetries of strong interaction
were the only instruments used for the description of this structure.\\
The models of hadrons (there are a lot of them) were not considered.
There was not also considered the structure of hadrons with heavy quarks.
The first round of problems (models of hadrons) is not considered
because of its unexhaustability. The second round of problems (hadrons
with heavy quarks) is not considered because of the physics of
heavy flavors tends mainly to electroweak interaction and will be
considered in the course of electroweak interactions.\\
Therefore, the presented here course, being the self-bounded part
of the theory of elementary particles, is considered separately.\\
These lectures were formed from 1989 and are the part of the set
of courses on the theory of fundamental interactions for the
students of Moscow Physical Technical Institute. The basic source of
the lectures was the unpublished manuscript "Hadrons and quarks" by
L.B. Okun. Another source was the "Lectures on the theory of unitary
symmetry in elementary particle physics" by Nguen Van Hieu (Atomizdat,
Moscow 1967). The list of references (due to its smallness) is absent.

\section{Lecture 1. Particles and interactions}
\subsection{The classes of particles and the types of interactions}

At present there exists a hard belief that all in the Nature is
constructed from elementary particles and that all the Natural
processes are caused
by the interactions of these particles.
Elementary particles are assumed today to be gauge
bosons,leptons,
quarks and Higgs scalar particles. The fundamental interactions
are assumed today to be strong,electroweak and gravitational interactions.
So, conventionally we can indicate
four classes of elementary particles
and three types of fundamental interactions.

The first class contains photon ($\gamma$), $W^+, W^-, Z$-bosons,
eight gluons $(g^a, a=1,...,8)$ and assumes the existence of graviton.
All these particles transfer the interactions: $\gamma, W^+, W^-, Z$-
bosons transfer electroweak interaction, gluons transfer strong
interaction and gravitons are hypothetical quanta of gravitational
field.

The second class contains leptons. At present we have six of them:
electron $(e)$, muon $(\mu )$, tau-lepton $(\tau )$ and corresponding
neutrino $(\nu_e ,\nu_{\mu}, \nu_{\tau})$. It is convenient to put six
leptons into three families
$$
\left(
\begin{array}{l}
\nu_e(<5.1 eV)\\e(0.511
MeV)\end{array}\right)
\left(
\begin{array}{c}
\nu_{\mu}(<0.27 MeV)\\ \mu(105.6
MeV)\end{array}\right)
\left(
\begin{array}{r}
\nu_{\tau}(<31 MeV)\\  \tau(1777
MeV)\end{array}\right)
$$

Neutrinos are electrically neutral; electron,muon and tau-lepton
possess electric charge. Leptons take part in electroweak and 
gravitational interactions.

The third class is represented by quarks. Today we know six quarks-
$ u, d, s, c, b, t$ and each quark is colored in one of three colors.
Like leptons quarks are represented by three families
$$
\left(
\begin{array}{l}u(2-8 MeV)\\d(5-15 MeV)\end{array}
\right)
\left(
\begin{array}{c}c(1-1.6 GeV)\\s(100-300 MeV)\end{array}
\right)
\left(
\begin{array}{r}t(175 GeV)\\b(4.1-4.5 GeV)\end{array}
\right)$$
Quarks are not seen as isolated particles. Together with gluons they
are the constituents of hadrons and there are some hundreds of hadrons.
Hadrons like quarks from which they are formed take part in all types
of interactions.

The forth class of particles are Higgs scalar particles not observed
yet experimentally. In minimal scheme of interactions there is one
Higgs scalar. The role of Higgs particles in the Nature is mainly
the theoretical one and is to make the electroweak interaction
renormalizable. In particular, the masses of all elementary particles
are due to the condensate of Higgs field. Probably, the existence of
Higgs fields is necessary for solving the fundamental problems
of cosmology: the homogeneity and causality of the Universe.

The following lectures on quark structure of hadrons will consider
light $u,d,s$-quarks and hadrons constructed from these quarks.
The main attention will be payed to classification of particles,
symmetries and conservation laws.

\subsection{Particles and antiparticles\\Fermions and bosons}

All particles have the partners-antiparticles that have the same
values of mass, spin, lifetime but have opposite in sign electric
charge, other charges, for instance, leptonic, baryonic, hypercharge,
strangeness etc. Antiparticle for electron $e^-$ is positron $e^+$,
for proton $p$ - antiproton $\bar p$, for neutron $n$ - antineutron
$\bar n$ etc. If the particle has no charges its antiparticle coincides
with the particle itself and the particle is called truly neutral.
The examples of truly neutral particles are $\gamma$-quantum, $Z^0$-
boson, $\pi^0$-meson etc. The part of the Universe surrounding us and
probably all the Universe is asymmetric: it consists of $e^-,n,p$ and
almost does not contain $e^+,\bar{p},\bar{n}$. The reasons of such
asymmetry are proposed in the theories of Grand Unification of particle
interactions.

All particles have either integer
or half-integer spin. The particles with half-integer spin are called
fermions and follow the Fermi-statistics according to which the given state
can be occupied by no more than one fermion. The wave function of the
system of fermions is antisymmetric under the intergange of fermion
variables. The particles with integer spin are called bosons and follow
the Bose-statistics according to which the given state
can be occupied by any number of bosons. The wave function of the
system of bosons is symmetric under the intergange of boson
variables. In what follows we will encounter the manifestations of
Fermi and Bose principles in the physics of hadrons many times.

\subsection{The characteristic of interactions in short}

Strong interaction is manifested on two levels: first, it is the interaction
of quarks inside hadrons. It is described by the gluon exchange at small
distances and by the confinement mechanism at large distances. Second,
it is the interaction of hadrons- of protons and neutrons in the nuclei,
for example. It is assumed that the second interaction is the result of
the first interaction although the exact mechanism of the realization of the
interaction between hadrons on the quark level is unclear. Hadrons are
divided on baryons (odd number of quarks and antiquarks, fermions) and
mesons (even number of quarks and antiquarks, bosons), on particles
(stable with respect to strong interaction, long lived) and resonances (decaying due to strong interaction, short lived). The division on particles 
and resonances becomes now more and more conventional one when short lived
decaying due to weak interaction hadrons are being discovered.

Electroweak interaction is mediated by the exchange of $\gamma,
W, Z$-particles. It splits into electromagnetic and weak interactions
at the energies lower than the masses of $W$ and $Z$-bosons. The massless
photon induces the long range interaction, the massive $W$ and $Z$-bosons
induce contact four-fermion interaction with dimensional Fermi constant.

Gravitational interaction of elementary particles is extremely  small
up to the energies of the order $10^{19} GeV$ where this interaction
could unify with other interactions.

\subsection{System of units}

Normally, in physics of elementary particles the velocities of the
order of light velocity $c$, the actions and angular momenta of
the order of Plank constant $\hbar$ are encountered. It is natural therefore
to use in physics of elementary particles the system of units where
$\hbar = c = 1$. Then, angular momentum, action and velocity are
dimensionless: $ [J]=[s]=[v]=1$. The energy and momentum have the
dimensionality of mass $[E]=[p]=[m]$, and time and length have the
dimensionality of inverse mass $[l]=[t]=[m^{-1}]$. As the unit of energy
it is convenient to choose $1 GeV$.

\subsection{Problem 1}

a) express $1 fm = 10^{-13}cm$ through $1 GeV^{-1}$,\\
b) evaluate the gravitational constant and find its value in the
system of units $\hbar=c=1$.
\newpage
\section{Ground states of light hadrons}
\subsection{Arithmetic of charges}

On the previous lecture we addressed to quarks. There are six quarks
$$
\left(
\begin{array}{l}u(5 MeV)\\d(7 MeV)\end{array}
\right)
\left(
\begin{array}{c}c(1.5 GeV)\\s(150 MeV)\end{array}
\right)
\left(
\begin{array}{r}t(175 GeV)\\b(4.5 GeV)\end{array}
\right)$$

If one compares the quark masses with the characteristic hadronic scale
$1 GeV$ it would be reasonable to split all quarks on light ones
($m_q << 1 GeV$) and heavy ones ($m_q > 1 GeV$). Below we will discuss
the properties of hadrons consisting of light quarks.

In strong and electromagnetic interactions the numbers of $u$-quarks,
$d$-quarks and $s$-quarks are conserved separately. Therefore we can
introduce the charges $U,D,S$ conserved in these interactions. Then
$$\begin{tabular}{cccc}
$\frac{charge}{quark}$ & $U$ & $D$ & $S$ \\
$u$ & $1$ & $0$ & $0$ \\
$d$ & $0$ &  $1$ & $0$ \\
$s$ & $0$ & $0$ & $1$
\end{tabular}
$$
Instead of $U,D,S$-charges it is convenient to introduce more familiar
charges that are linear combinations of $U,D,S$
$$B = \frac{U+D+S}{3}$$ (baryon number),
$$T_3 = \frac{U-D}{2}$$ (3-d component of isotopic spin),
$$Y = \frac{U+D-2S}{3}$$ (hypercharge).
Then
$$
\begin
{tabular}{ccccc}
$\frac{charge}{quark}$&$B$&$T_3$&$Y$&$Q$\\
$u$&1/3&1/2&1/3&2/3\\
$d$&1/3&-1/2&1/3&-1/3\\
$s$&1/3&0&-2/3&-1/3
\end{tabular}
$$
Baryonic charge is defined in such a way that the proton (or any other
baryon) consisting of three quarks has baryonic charge equal to one.
Baryonic charge is conserved in all interactions. 3-d component of
isotopic spin is defined in such a way as $u$ and $d$ quarks form the
isotopic doublet ($T=1/2$) and strange quark is isosinglet. Finally,
hypercharge is defined in such a way  that there holds the relation
of Gell-Mann - Nishigima
$$Q = T_3 +\frac{Y}{2}~.$$

In strong and electromagnetic interactions both $T_3$ and $Y$ are
conserved. In weak interactions, where the transitions
$d \rightarrow u (\Delta T_3 = 1, \Delta Y = 0)$ and
$s \rightarrow u (\Delta T_3 = 1/2, \Delta Y = 1)$ are possible $T_3$ and $Y$
are not conserved but their violation has very definite character.

Strong interactions of quarks are due to their color charges and
these charges are identical for $u,d,s$- quarks. Therefore strong
interactions of $u,d,s$- quarks are identical. If the quark masses
are equal $m_u = m_d = m_s$ then $u,d,s$- quarks will be indistinguishable
by strong interactions i.e. it could be possible to change $u$-quark
by $d$-quark or by $s$-quark in hadrons and neither masses nor
strong interaction of these hadrons would be changed. In reality
$m_u \neq m_d \neq m_s$ but the inequalities look small compared
to the characteristic hadronic scale $1 GeV$, especially small 
in the case of $u$ and
$d$-quarks. Therefore, it is possible to expect an almost
exact $SU(2)$-symmetry (isotopic spin)
$$\left(
\begin{array}{l}u\\ d\end{array}\right)
 \rightarrow U(2)
\left(
\begin{array}{r}u\\ d\end{array}\right)
$$
and approximate $SU(3)$-symmetry
$$\left(
\begin{array}{l}u\\d \\s\end{array}\right)
\rightarrow U(3)
\left(
\begin{array}{r}u\\ d\\s\end{array}\right)
$$

The results of the symmetries are the conserved charges-the generators
of these symmetries. In the case of $SU(2)$-symmetry the conserved
charges are presented by isotopic spin ${\bf T}$. Note that $T_3$ is
the diagonal generator of $SU(2)$-group and $T_3$ and $Y$ are the
diagonal generators of $SU(3)$-group. The states of particles (hadrons)
are classified by irreducible representations of these groups,
in particular, for $SU(2)$-group the irreducible representations
are labeled by the square of isotopic spin ${\bf T}^2 = T(T+1)$
where $T = {T_3}_{max}$ in the multiplet.

Let us classify the ground states of mesons and baryons by
$T, T_3$ and $Y$.

\subsection{Mesons}

Basic states of mesons are presented by 9  $q{\bar q}$ mesons
with spin $0$ ($1$). They are:\\
isotopic triplet of $\pi(140) (\rho(770))$-mesons
$$T=1; T_3 = +1,0,-1; Y=0$$
$$
\left(
\begin{array}{l}u{\bar d}\\
\frac{1}{\sqrt{2}}(u{\bar u}-d{\bar d})\\
d{\bar u}
\end{array}
\right)
=
\left(
\begin{array}{c}
\pi^+\\
\pi^0\\
\pi^-\end{array}\right)
,
\left(
\begin{array}{r}
\rho^+\\
\rho^0\\
\rho^-\end{array}\right);
$$
isotopic doublets of $K (490)(K^*(890))$ - mesons
$$T=1/2; T_3 = +1/2,-1/2; Y=+1,-1$$
$$
\left(
\begin{array}{l}u{\bar s}\\
d{\bar s}\end{array}\right)=
\left(\begin{array}{c}K^+\\K^0\end{array}\right),
\left(\begin{array}{r}K^{+*}\\K^{0*}\end{array}\right);$$
$$
\left(
\begin{array}{l}s{\bar d}\\
s{\bar u}\end{array}\right)=
\left(\begin{array}{c}{\bar K}^0\\K^-\end{array}\right),
\left(\begin{array}{r}{\bar K}^{0*}\\K^{-*}\end{array}\right);$$
isotopic singlets $\eta, \eta^\prime$ ($T=0,T_3=0,Y=0$)
$$\frac{1}{\sqrt(6)}(u{\bar u}+d{\bar d}-2s{\bar
s})=\eta(550)$$
$$\frac{1}{\sqrt(3)}(u{\bar u}+d{\bar d}+s{\bar
s})=\eta^\prime(960)$$
for scalar particles \\and isotopic singlets
$\omega, \phi$ ($T=0,T_3=0,Y=0$)
$$\frac{1}{\sqrt(2)}(u{\bar u}+d{\bar d})=\omega(780)$$
$$s{\bar s}=\phi(1020)$$
for vector particles.

\subsection{Baryons}

Out of 27=3*3*3 flavor states of $qqq$-baryons with spin
$1/2$ and $3/2$ the Fermi statistics admits only the octet
of baryons with spin $1/2$ ((8,1/2)) and decuplet of baryons
with spin $3/2$  ((10,3/2)). The octet of baryons is presented by:
isotopic triplet of $\Sigma(1190)$ - hyperons
$$T=1; T_3 = +1,0,-1; Y=0$$
$$\left(\begin{array}{l}uus\\uds\\dds\end{array}\right)=
\left(\begin{array}{r}\Sigma^+\\ \Sigma^0\\
\Sigma^-\end{array}\right);$$
isotopic doublets of nucleons $(N(939))$ and $\Xi(1320)$ - hyperons
$$T=1/2; T_3 = +1/2,-1/2; Y=+1,-1$$
$$\left(\begin{array}{l}uud\\ddu\end{array}\right)=
\left(\begin{array}{r}p\\n\end{array}\right);$$
$$\left(\begin{array}{l}ssu\\ssd\end{array}\right)=
\left(\begin{array}{r}\Xi^0\\\Xi^-\end{array}\right);$$
isotopic singlet ($T_3=0,Y=0$)- $\Lambda(1116)$ - hyperon
consisting of $uds$-quarks.

The decuplet of baryons with spin $3/2$ is presented by:
quartet of isobars $\Delta(1232)$
$$T = 3/2; T_3 = +3/2,+1/2,-1/2,-3/2; Y = 1$$
$$\left(\begin{array}{l}uuu\\uud\\ddu\\ddd\end{array}\right)=
\left(\begin{array}{r}\Delta^{++}\\\Delta^{+}\\\Delta^{0}\\\Delta^{-}
\end{array}\right);$$
isotopic triplet of $\Sigma^*(1385)$ - hyperons
$$T=1; T_3 = +1,0,-1; Y=0$$
$$\left(\begin{array}{l}uus\\uds\\dds\end{array}\right)=
\left(\begin{array}{r}\Sigma^{*+}\\\Sigma^{*0} \\\Sigma^{*-}
\end{array}\right);$$
isotopic doublet of $\Xi^{*}(1530)$ - hyperons
$$T=1/2; T_3 = +1/2,-1/2; Y=-1$$
$$\left(\begin{array}{l}ssu\\ssd\end{array}\right)=
\left(\begin{array}{r}\Xi^{*0}\\\Xi^{*-}\end{array}\right)$$
and isotopic singlet ($T_3=0,Y=-2$)- $\Omega^-(1672)$ - hyperon
consisting of $sss$-quarks.

\subsection{Decays}

The lightest out of all hadrons $\pi^0$ - meson is decaying
due to electromagnetic interaction to two $\gamma$ - quanta
$$\pi^0 \rightarrow 2\gamma$$
and lives about $10^{-16} $ seconds.\\
Its charged partners $\pi^{\pm}$ are decaying due to weak interaction
$$\pi^{\pm} \rightarrow \mu^{\pm}\nu_\mu ({\bar
\nu}_{\mu})$$
and live about $10^{-8}$ seconds.\\
$K$ -mesons are decaying due to weak interaction to
$2\pi, 3\pi, ...$  and have lifetimes from $10^{-8}$ to $10^{-10}$
seconds.\\
$\eta$ - meson is decaying to $3\pi
, 2\gamma, ...$ violating the isotopic symmetry. Its decay to
$2\pi$ - mesons is forbidden by $CP$ - symmetry (see lecture 3 below).\\
Meson $\eta^\prime$ is decaying to $\eta \pi \pi$ due to
strong interaction.\\
Proton is stable. At present the lower bound of its lifetime is $10^{32}$ years.
Neutron has lifetime 887 seconds and decays to proton, electron and
electron antineutrino. $\Lambda, \Sigma^{\pm}, \Xi^0,
\Xi^-, \Omega^-$ - hyperons are decaying due to weak interaction,
$\Sigma^0$ - hyperon is decaying electromagnetically to $\Lambda \gamma$
and $\Delta, \Sigma^* $, $\Xi^*$ - particles are resonances i.e.
are decaying due to strong interactions.

\subsection{Problem 2}

a) Why the decay $\Xi \rightarrow \Sigma \pi$ is not seen
experimentally?\\
b) How has the hypercharge $Y$ to look like in order to
formula $Q = T_3 + \frac{Y}{2}$  be valid for all hadrons
including the hadrons with heavy ($c,b,t$) - quarks?

\section{Lecture 3. Discrete symmetries}

In the present lecture we will turn to the discrete symmetries-
space inversion, charge conjugation and G - transformation.
These symmetries will help us to understand the peculiarities
of the decays of $\eta, \eta^\prime, \rho, \omega, \phi$ - mesons.

\subsection{Space inversion}

The transformation of space inversion is realized by the change
 of the reference frame from $t,x,y,z$ to $t,-x,-y,-z$. Various
objects transform under this change of the  reference frame
differently. So, the polar vectors of space position and
momentum $({\bf r,p})$ change their signs, the axial vector
of angular momentum $({\bf l},
l_i=\epsilon_{ijk}r_ip_k)$ is unchanged. The scalar product of two
polar vectors forms the scalar that is also unchanged under
the transformation of space inversion  but the scalar product of
polar and axial vectors forms the pseudoscalar that change the
sign under  space inversion . The spatial parity ($P$) of some
object is determined by its property to change or not to change
the sign under space inversion . Scalar and axial vector are
$P$ - even but pseudoscalar and vector are $P$ - odd.

Now we know that the Lagrangians of strong and electromagnetic
interactions are not changed under the transformation of space inversion
i.e. are the scalars but the Lagrangian of weak interaction is
the sum of scalar and pseudoscalar. Invariance
with respect to space inversion on the quantum-mechanical
level means that the amplitude of the process and the
amplitude of the space inverted process coincide
$$<b^P|V|a^P> = <b|P^{-1}VP|a> = <b|V|a>.$$
If the initial and final states have definite $P$ - parity,
$$P|a> = p_a|a>$$
$$P|b> = p_b|b>,$$
then $p_a\cdot p_b = 1$ or $p_a = p_b$ i.e. $P$ - parity
is conserved.

{\bf P - parities of the particles}\\
Strong and electromagnetic interactions that conserve the
$P$ - parity conserve also $U,D,S$ - charges i.e. the numbers
of $u,d,s$ - quarks.\\
Therefore, the quark parities $p_u,p_d,p_s$ can be chosen
arbitrarily.
It is accepted to take them equal to 1: $p_u = p_d = p_s = 1$.
One of the theorem of the theory of fermions based on the
Dirac equation is the statement that the product of the
internal parities of fermion and antifermion is equal to -1:
$p_up_{\bar u} = p_dp_{\bar d} = p_sp_{\bar s}
= -1$ . This means that $p_{\bar u} = p_{\bar d} =
p_{\bar s} = -1$ for  our convention $p_u = p_d = p_s = 1$.
The internal parity of the system of two particles (of the
meson consisting from quark and antiquark, for example)
is equal to the product of the internal parities of the
constituent particles and of the orbital parity of their
relative motion
$$\Psi ({\bf r}_q - {\bf r}_{\bar q}) \begin{array}{c} P\\\rightarrow
\end{array} p_qp_{\bar q} \Psi ({\bf r}_{\bar q} - {\bf r}_q) =
p_qp_{\bar q} (-1)^l\Psi ({\bf r}_{q} - {\bf r}_{\bar q}) ,$$
i.e.
$$P(M(q{\bar q})) = (-1)^{l+1},$$
where $l$ is the orbital momentum of relative motion of quarks.
For ground states of mesons $l = 0$ and $P(M(q{\bar q})) = -1$
(pseudoscalar and vector mesons).\\
The internal parity of the system of three particles (of the baryon
consisting from three quarks, for example)
is equal to the product of the internal parities of the
constituents, the parity of relative orbital motion of
any pair of constituents and the parity of relative orbital motion 
of chosen pair and third constituent
$$ p(qqq) = p_qp_qp_q (-1)^l(-1)^L,$$
where $l$ is the orbital momentum of the pair, $L$ is the relative
orbital momentum of the pair and third particle . For ground
states of baryons $l = L = 0$ and  $P(B(qqq)) =
1$.

\subsection{Charge conjugation}

Charge conjugation, by definition, transforms particle to
antiparticle, all the charges of which have different sign
with respect to the charges of particle. Truly neutral particles
(not only electrically neutral) under $C$ - conjugation transform
to themselves what allows to introduce the notion of $C$ - parity.
For example, wave function (or the state) of some particles 
($ \gamma, \rho^0, \omega, \phi$)
changes the sign under charge conjugation
$$ C\gamma = - \gamma, C_{\gamma} = -1, ... ,$$
wave function of other particles ($\eta, \eta^\prime, \pi^0$)
does not change the sign under charge conjugation
$$C\eta = \eta, C_\eta = 1, ... .$$

$C$ - parity of mesons constructed from quark and antiquark
is determined by the relative orbital momentum of quark and antiquark
and their total spin $s$
$$\begin{array}{l}\\
\psi({\bf r}_q - {\bf r}_{\bar q})\psi(s_q,s_{\bar q})a_q^+(
{\bf r}_q,s_q)a_{\bar q}^+({\bf r}_{\bar q},s_{\bar q})|0>
\end{array}
\begin{array}{c} C\\\rightarrow \end{array}
\begin{array}{r}\\
\psi({\bf r}_q - {\bf r}_{\bar q})\psi(s_q,s_{\bar q})a_{\bar q}^+(
{\bf r}_q,s_q)a_{q}^+({\bf r}_{\bar q},s_{\bar q})|0>
\end{array}$$
$$= -
\psi({\bf r}_q - {\bf r}_{\bar q})\psi(s_q,s_{\bar q})
a_{q}^+({\bf r}_{\bar q},s_{\bar q})a_{\bar q}^+(
{\bf r}_q,s_q)|0>$$
$$= (-1)^{1+l+s+1}
\psi({\bf r}_q - {\bf r}_{\bar q})\psi(s_q,s_{\bar q})a_q^+(
{\bf r}_q,s_q)a_{\bar q}^+({\bf r}_{\bar q},s_{\bar q})|0>
$$
(the summation over the variables ${\bf r}_q, {\bf r}_{\bar q}, s_q,
s_{\bar q}$ is here assumed). In the above transformations we have
taken into account that fermion operators anticommute, that the
intergange of quark and antiquark coordinates results in the
multiplier $(-1)^l$ and that the intergange of quark and antiquark
spins gives multiplier $(-1)^{s+1}$. So, $C$ - parity of meson
$M = q{\bar q}$  is equal to
$$ C(M(q{\bar q})) = (-1)^{l+s},$$
i.e. is negative for vector mesons ($l=0,s=1; \rho^0,\omega,\phi$)
and positive for pseudoscalar mesons ($l=0,s=0; \pi^0,
\eta,\eta^\prime$).

\subsection{$G$ - transformation}

Under the charge conjugation the components of isotopic triplet
of $\pi$ - mesons transform as follows
$$\begin{array}{l}\\\pi^0\end{array}~~~~~~  \begin{array}{c} C\\
\rightarrow
\end{array}~~~~~~\begin{array}{c}\\\pi^0\end{array}~~~~~~~~~
\begin{array}{c}\\\pi^\pm \end{array}~~~~~~\begin{array}{c} C\\
\rightarrow \end{array} ~~~~~~\begin{array}{r}\\\pi^\mp\end{array}.$$
It can be seen that charged $\pi$ - mesons are not the eigenstates
of charge conjugation operator, so, no $C$ - parity can be ascribed
to them. It is possible however to join charge conjugation
with some rotation in isotopic space in order to both neutral and charged
$\pi$ - mesons be eigenstates with respect to
the combined transformation ($G$ - transformation).
The resulting $G$ - parity is conserved in strong interaction
in the same degree as the charge conjugation and isotopic
rotations are the symmetries of strong interaction. The latter
(the symmetry of isotopic rotations) is violated only slightly
by the difference of $u$ and $d$ - quark masses, so, $G$ - transformation
is almost an exact symmetry of  strong interaction.\\
The desired rotation in isotopic space  is the rotation around
the 2-nd axis on the angle $180^o $
$$\begin{array}{l}\\\pi^0\end{array}~~~~~~  \begin{array}{c}
T_2^{180}\\ \rightarrow
\end{array}~~~~~~\begin{array}{c}\\-\pi^0\end{array}~~~~~~~~~
\begin{array}{c}\\\pi^\pm \end{array}~~~~~~\begin{array}{c}
T_2^{180}\\ \rightarrow \end{array}
~~~~~~\begin{array}{r}\\-\pi^\mp\end{array}$$
Then
$$G:\pi=CT_2^{180}:\pi = -\pi,$$
i.e. $G_{\pi} = -1$. For other particles $G$ -parity can
be easily determined
$$G_\eta = +1,~~~~~~G_\rho = +1,~~~~~~G_\omega = -1,~~~~~~G_\phi =
-1.$$

\subsection{Decays}

Let us apply the discussed above discrete symmetries to the decays
of $\eta, \rho, \omega$ - mesons. The following table of
forbidden by appropriate symmetry decays is obvious
$$\begin{array}{l}\eta \rightarrow \pi\pi\\
\eta \rightarrow \pi\pi\pi\\
\eta \rightarrow \pi^0\gamma\\
\eta \rightarrow \pi^0\pi^0\gamma\\
\rho \rightarrow \pi^0\pi^0\\
\rho \rightarrow \pi\pi\pi\\
\omega \rightarrow \pi\pi\\
\omega \rightarrow \pi^0\pi^0\pi^0\end{array}
\begin{array}{r}P,CP\\G\\C\\C\\C,Bose\\G\\G\\C\end{array}$$

\subsection{Problem 3}

Write allowed and forbidden quantum numbers $J^{PC}$  for
$q{\bar q}$ - system.

\newpage
\section{Lecture 4. Isotopic symmetry}
\subsection{Quarks and antiquarks}

Isotopic symmetry of strong interaction is the result of
the approximate equality of $u$ and $d$ - quark masses
($m_u = 4 MeV, m_d = 7 MeV, m_u \approx m_d$) from the point
of view of characteristic hadronic scale ($1 GeV$)
$$\Delta m_{ud} << 1 GeV$$
and is manifested in the invariance of strong interaction for
$m_u = m_d$ under isotopic $SU(2)$ - transformations
$$\left(
\begin{array}{l}u\\ d\end{array}\right)
 \rightarrow U(2)
\left(
\begin{array}{r}u\\ d\end{array}\right).
$$
The lowest (fundamental) representation of $SU(2)$ - group is
the spinor $\Psi^\alpha; \alpha = 1,2$, the doublet of quarks,
for example. The $SU(2)$ - group
$$U^+U = I~~~~~~~~~~~~~~~~~~~~~~~~det U = 1$$
has three generators $\frac{ \tau_i}{2}$
$$U = exp(i {\vf \omega}\frac{{\vf \tau}}{2}),$$
one of which ($\frac{\tau_3}{2}$) can be made diagonal. The
matrices $ \tau_i$ are hermitian (unitarity of matrix $U$), 
traceless (the consequence of the constraint $det U = 1$)
and have the form
$$\tau_1 = \left(\begin{array}{l}
0~~~~1\\1~~~~0\end{array}\right) \tau_2 = \left(\begin{array}{c}
0-i\\i~~~~0\end{array}\right) \tau_3 = \left(\begin{array}{r}
1~~~~0\\0-1\end{array}\right).$$
They have the following commutation, anticommutation and
normalization relations
$$\left[\frac{\tau_i}{2},\frac{\tau_j}{2}\right] =
i\epsilon_{ijk}\frac{\tau_k}{2},$$
$$\left\{\tau_i ,\tau_j\right\} = 2\delta_{ij}I,$$
$$Tr(\tau_i\tau_j) = 2\delta_{ij}.$$
For other representations of  $SU(2)$ - group the  $SU(2)$ -
 transformations have similar structure
 $$exp(i{\vf \omega}{\vf T}),$$
$$\left[T_i ,  T_j\right] =
i\epsilon_{ijk}T_k .$$
The quark doublet is transformed like contravariant spinor
$$q^{\prime\alpha} = {U^\alpha}_\beta q^\beta~.$$
Antiquarks are transformed by complex conjugate matrix $U^*$
$$q^{\prime\alpha *} = {U^{\alpha *}}_\beta q^{\beta *}$$
or like covariant spinor
$${\bar q}^{\prime}_\alpha = {\bar q}_\beta {U^{-1\beta}}_\alpha =
 {\bar q}_\beta {U^{+\beta}}_\alpha = {\bar q}_\beta
 {U^{*\alpha}}_\beta.$$
 The $SU(2)$ - group has two invariant tensors
 $${\delta^\alpha}_\beta \rightarrow {U^\alpha}_{\alpha^\prime}
{\delta^{\alpha^\prime}}_{\beta^\prime} {U^{-1{\beta^\prime}}}_\beta
= {\delta^\alpha}_\beta,$$
$$\epsilon^{\alpha\beta} \rightarrow {U^\alpha}_{\alpha^\prime}
{U^\beta}_{\beta^\prime}\epsilon^{\alpha^\prime\beta^\prime} =
 \epsilon^{\alpha\beta} (det U) = \epsilon^{\alpha\beta}.$$
 Using the second invariant tensor $\epsilon^{\alpha\beta}
$ the covariant spinor can be transformed to the contravariant one
i.e. quarks and antiquarks are transformed as unitary equivalent
representations
$${\bar q}_\alpha = ({\bar u},{\bar d})\begin{array}{c}\epsilon\\ \\ \
\rightarrow \end{array}\left(\begin{array}{c}{\bar d}\\-{\bar
u}\end{array}\right) = {\bar q}^\beta = \epsilon^{\beta\alpha}
{\bar q}_\alpha.$$

\subsection{Diquarks and mesons}

The state space (isotopic) of two quarks has four basis elements
$q^\alpha q^\beta$. Isotopic wave functions of diquarks, the tensors
$Q^{\alpha\beta}$, form the state space that is reducible with respect
to transformations of $SU(2)$ - group. It can be divided on two invariant
(irreducible) subspaces: $S^{\alpha\beta}$ - subspace of symmetric tensors
and $A^{\alpha
\beta} = \epsilon^{\alpha\beta} A$ - subspace of antisymmetric tensors.
The subspace $S^{\alpha\beta}$ has three basis elements:
$uu, \frac{1}{\sqrt{2}}(uu+dd),dd$. They form triplet ($T = 1$) where
$T_3 = 1,0,-1$, correspondingly. The subspace $A^{\alpha\beta}$ is
one dimensional with one basis element $\frac{1}{\sqrt{2}}(uu-dd)$.
It represent an isotopic singlet ($T = 0$).

Analogously, in the isotopic  space of states of quark and antiquark
(mesons) we have the basis elements $q^\alpha{\bar q}_\beta$ and
isotopic wave functions ${Q^\alpha}_\beta =
{Q^\alpha_0}_\beta + {\delta^\alpha}_\beta Q ({Q^\alpha_0}_\alpha =
0)$. Three basis elements of ${Q^\alpha_0}_\beta$ - subspace
 $(u{\bar d}, \frac{1}{\sqrt{2}}(u{\bar u} - d{\bar d}), d{\bar u})$ 
 form isotopic triplet and one basis element of ${\delta^\alpha}_\beta Q$ - 
 subspace ($\frac{1}{\sqrt{2}}(u{\bar u} +
  d{\bar d}) $) forms isotopic singlet. Let us write (having in mind
pseudoscalar mesons) the triplet ${Q^\alpha_0}_\beta$ as follows
 $${Q^\alpha_0}_\beta = \left(\begin{array}{c}
\frac{\pi_0}{\sqrt{2}}~~~~\pi^+\\ \pi^- -\frac{\pi_0}{\sqrt{2}}
\end{array}\right)$$ and compare this presentation of tensor
${Q^\alpha_0}_\beta$ with another one where the isotopic vector
${\vf \pi}$ is explicitly introduced
$${Q^\alpha_0}_\beta = \frac{1}{\sqrt{2}}{\vf
\pi}{{\vf \tau}^\alpha}_\beta = \left(\begin{array}{c}
\frac{\pi_3}{\sqrt{2}}~~~~~~~~~~\frac{1}{\sqrt{2}}(\pi_1+i\pi_2)\\
\frac{1}{\sqrt{2}}(\pi_1-i\pi_2)~~~~~~-\frac{\pi_3}{\sqrt{2}}
\end{array}\right).$$
As a result we have the following one to one correspondence
between the two forms of the representation of isotopic triplet
$$\pi^\pm = \frac{1}{\sqrt{2}}(\pi_1\pm i\pi_2), \pi_0 = \pi_3,$$
that was already use in the preceding lecture.

\subsection{Isotopic states in the systems of two and three pions}

Two pions ${\vf \pi}_1,{\vf \pi}_2$ can have the total isospin equal
to $0,1,2$ $(3*3 = 1+3+5)$. An explicit construction of the isotopic
wave function of the system of two pions
$$ T=0~~~~~~~~~~~{\vf \pi}_1\cdot{\vf \pi}_2 = \delta_{ij}\pi_{1i}
\pi_{2j}~~~~~~~~~~~S$$
$$ T=1~~~~~~~~~~~{({\vf \pi}_1\times{\vf \pi}_2)}_k =
\epsilon_{ijk}\pi_{1i} \pi_{2j}~~~~~~~~~~~A$$
$$ T=2 ~~~~~~~~~~~\pi_{1i}\pi_{2j}+\pi_{1j}\pi_{2i}-
\frac23\delta_{ij}\pi_{1k}
\pi_{2k}~~~~~~~~~~~S$$
shows that the isotopic wave function of the system of two pions
is symmetric ($S$) for the total isospin $T=0,2$ and antisymmetric
($A$) for $T=1$. Therefore, in accordance with Bose principle
the orbital angular momentum of two pions ($l$) can be even for
$T=0,2$ and odd for $T=1$. So, $\omega $ - meson ($T=0,S=1$)
cannot decay to two pions due to the isospin conserving part of strong 
interaction. In the previous lecture we have seen that the decay
$\omega \rightarrow 2\pi$ is forbidden by the conservation of
$G$ - parity, now we have looked at this problem from another side.

For the system of three pions ${\vf \pi}_1,{\vf \pi}_2,{\vf \pi}_3$
where $3*3*3 = (1+3+5)*3 = 3+(1+3+5)+(3+5+7)$ we have one singlet
$${\vf \pi}_1\cdot{\vf \pi}_2\times{\vf \pi}_3 = \epsilon_{ijk}
\pi_{1i}\pi_{2j}\pi_{3k}~~~~~~~~~~~~~~~~~A~,$$
three triplets one of which can be chosen with symmetric wave function
$${\vf \pi}_1({\vf \pi}_2\cdot{\vf \pi}_3)+
{\vf \pi}_2({\vf \pi}_3\cdot{\vf \pi}_1)+
{\vf \pi}_3({\vf \pi}_1\cdot{\vf \pi}_2) ~~~~~~~~~~~~S$$
etc. If one assumes the conservation of isotopic spin in the
decay $\eta \rightarrow 3\pi$ ($T=0$) then in accordance with Bose principle
the coordinate/momentum wave function of three pions should be
antisymmetric. This leads to the negative $C$ - parity of the
system of three pions and to the violation of $C$ - parity  in the decay.
If one assumes that the coordinate/momentum wave function of three pions
in the considered decay is symmetric (there is no relative orbital
motion of pions due to small phase space, for example) 
then $C$ - parity is conserved
 but the total isospin
of pions cannot equal to zero i.e. isospin is not conserved.
Before (see lecture 3) we have seen that $G=CT_2^{180}$ - parity is violated
in $\eta \rightarrow 3\pi$ decay, the presented here consideration
so to say disentangle 
this violation.

\subsection{Pion-nucleon scattering}

Let us apply the isotopic symmetry to the pion-nucleon scattering.
There are 10 processes
$$\begin{array}{l}
\pi^+p\rightarrow \pi^+p\\
\pi^0p\rightarrow \pi^0p\\
\pi^-p\rightarrow \pi^-p\\
\pi^0p\rightarrow \pi^+n\\
\pi^-p\rightarrow \pi^0n
\end{array}
\begin{array}{c}T_2^{180}\\\rightarrow\end{array}
\begin{array}{r}
\pi^-n\rightarrow \pi^-n\\
\pi^0n\rightarrow \pi^0n\\
\pi^+n\rightarrow \pi^+n\\
\pi^0n\rightarrow \pi^-p\\
\pi^+n\rightarrow \pi^0p~,
\end{array}            $$
where the processes in the right column are isotopically
connected to the processes in the left column by the $180^o$ rotation
around the 2-nd axis in isotopic space. So, we have to consider
only the processes in the left column. The fourth process of
the left column is time reverse of the fifth process of the right column,
so, its amplitude is equal to the amplitude of the fifth process
(no matter of which column). The second process is not observed
experimentally because of the lack of $\pi^0$ - beams. The remaining
three processes $\pi^+p\rightarrow \pi^+p,\pi^-p\rightarrow \pi^-p,
\pi^-p\rightarrow \pi^0n$ have the amplitudes $M_+,M_-,M_0$
connected by isotopic symmetry. Considering the states of
pion and nucleon with definite total isospin one gets
$$\pi^+p = (\frac32,+\frac32)$$ $$\pi^-p =
\sqrt{\frac13}(\frac32,-\frac12)-\sqrt{\frac23}(\frac12,-\frac12)$$
$$\pi^0n =
\sqrt{\frac23}(\frac32,-\frac12)+\sqrt{\frac13}(\frac12,-\frac12).$$
Due to isotopic symmetry the amplitude of pion-nucleon
scattering does not depend on the value of third component of
isospin i.e. we have only two independent amplitudes $M_{3/2}$ and
$M_{1/2}$.  Thus $M_+,M_-,M_0$ are expressed through $M_{3/2}$ and
$M_{1/2}$
$$ M_+ = M_{3/2}$$
$$ M_- = \frac13 M_{3/2} + \frac23 M_{12}$$
$$ M_0 = \frac{\sqrt{2}}{3}(M_{3/2}-M_{1/2})$$
and 
$$ \sqrt{2} M_0 + M_- = M_+.$$ The last relation is just what we
wanted to obtain. For the energy of incoming pion corresponding
to the $\Delta$ - resonance region one amplitude $M_{3/2}$ is
much larger than the other $M_{1/2}$. In that case
$$M_+:M_-:M_0 \approx 1:\frac13:\frac{\sqrt{2}}{3},$$
and the cross sections follow the ratios
$$\sigma_+:\sigma_-:\sigma_0 \approx 9:1:2.$$
These ratios are in good agreement with experiment.

\subsection{Problem 4}

Find the ratio of the widths of $\eta \rightarrow \pi^0\pi^0\pi^0$
and $\eta \rightarrow \pi^+\pi^-\pi^0$ decays assuming that
the total isospin of three pions is equal to $T=1$ and that the
coordinate/momentum wave function of pions is symmetric
(see lecture).

\newpage
\section{$SU(3)$ - symmetry}
\subsection{Fundamental representation}

The comparison of $u,d,s$ - quark masses with characteristic
hadronic scale ($m_u = 4 MeV, m_d = 7 MeV, m_s = 150 MeV << 1 GeV$)
tells us that $SU(3)$ - symmetry
$$\left(
\begin{array}{l}u\\ d\\s\end{array}\right)
 \rightarrow U(3)
\left(
\begin{array}{r}u\\ d\\s\end{array}\right).
$$
can be considered as an approximate symmetry of strong interaction.
Mainly it is violated by the mass difference of strange ($s$) and
nonstrange ($u$ and $d$) quarks. The consequences of just this
reason of violation of $SU(3)$ - symmetry are the mass formulae
for ground states of mesons and baryons that will be considered
in the next lecture. In the present lecture we will classify the
$SU(3)$ - multiplets of mesons and baryons.

The fundamental representation of $SU(3)$ - group is the triplet
of quarks
$$
q^\alpha =
\left(
\begin{array}{r}u\\ d\\s\end{array}\right).
$$
The $SU(3)$ - transformations of the triplet are produced by
the unitary matrices 3$\times$3
$$U = exp(i { \omega^a}\frac{{ \lambda^a}}{2}),$$
where $\lambda^a$  are Gell-Mann matrices
$$\lambda^{1,2,3} = \left(\begin{array}{l} \tau^{1,2,3}~~~~0\\
                                                0~~~~~~~~~0
\end{array}\right); \lambda^{4,5} = \left(
\begin{array}{r}     0~~~~~0~~1(-i)\\
                     0~~~~~~0~~~~~~0\\
                     1(+i)~~0~~~~~~0\end{array}\right)$$
$$                    \lambda^{6,7} = \left(
\begin{array}{l}
                     0~~~~~~0~~~~~~0\\0~~~~~0~~1(-i)\\
                     0~~1(+i)~~~~0\end{array}\right);
\lambda^8 = \frac{1}{\sqrt{3}}\left(\begin{array}{r}
1~~~~~~0~~~~~~0\\
0~~~~~~1~~~~~~0\\
0~~~~~~0~~~-2\end{array}\right).$$
The $SU(3)$ - group has two diagonal generators: the third
component of isotopic spin and hypercharge,
$$T_3 = \frac{\lambda^3}{2} ~~~~~~~~~Y = \frac{1}{\sqrt{3}}
\lambda^8,$$
that are conserved in strong and electromagnetic interactions
(see lecture 2).

Antiquarks are transformed as complex conjugate of the fundamental
(contravariant spinor) representation and these transformations
coincide with the transformations of the covariant spinor due to the
unitarity of the group. Unlike the case of $SU(2)$ - group the
covariant spinor representation is not now equivalent to the
contravariant representation (quarks and antiquarks transform
nonequivalently).

The $SU(3)$ - group has two invariant tensors
$${\delta^\alpha}_\beta \rightarrow {U^\alpha}_{\alpha^\prime}
{\delta^{\alpha^\prime}}_{\beta^\prime} {U^{-1{\beta^\prime}}}_\beta
= {\delta^\alpha}_\beta,$$
$$\epsilon^{\alpha\beta\gamma} \rightarrow {U^\alpha}_{\alpha^\prime}
{U^\beta}_{\beta^\prime}{U^\gamma}_{\gamma^\prime}
\epsilon^{\alpha^\prime\beta^\prime\gamma^\prime} =
 \epsilon^{\alpha\beta\gamma} (det U) =
 \epsilon^{\alpha\beta\gamma}.$$
 
\subsection{Mesons}

Mesons consist of quark and antiquark and are described by
the tensor (flavor wave function)
$${Q^\alpha}_\beta =
{Q^\alpha_0}_\beta + \frac13{\delta^\alpha}_\beta Q
~~~~~~~~~~~~({Q^\alpha_0}_\alpha = 0),$$
two parts of which represent the octet and singlet states.
The basis states of the octet are (we indicate pseudoscalar mesons
as an example):\\
isotopic triplet of $\pi$ - mesons
$$u{\bar d}, \frac{1}{\sqrt{2}}(u{\bar u} - d{\bar d}), d{\bar u};$$
isotopic doublets of $K$ - mesons
$$u{\bar s},d{\bar s}; ~~~~~~~~~~~s{\bar d}, s{\bar u}$$
and isotopic singlet $\eta_8$
$$\frac{1}{\sqrt{6}}(u{\bar u} + d{\bar d} - 2s{\bar s}).$$
The basis state of the singlet is $\eta_0$
$$\eta_0 = \frac{1}{\sqrt{3}}(u{\bar u} + d{\bar d} + s{\bar s}).$$

Using the above octet basis states the octet part of tensor 
${Q^\alpha}_\beta$ (${Q^\alpha_0}_\beta$) can be represented
in the following form
$${Q^\alpha_0}_\beta = \left(\begin{tabular}{ccc}
$\frac{\pi_0}{\sqrt{2}}+\frac{\eta_8}{\sqrt{6}}$&$\pi^+$&$K^+$\\
$\pi^-$ &$ -\frac{\pi_0}{\sqrt{2}}+\frac{\eta_8}{\sqrt{6}}$&$K^0$\\
$K^-$&${\bar K}^0$&$-\frac{2\eta_8}{\sqrt{6}}$ \end{tabular}\right)~.$$

\subsection{Baryons}

Baryons consist of three quarks. Their wave function presented
by the tensor $B^{\alpha\beta\gamma}$ is reducible. Let us briefly
describe the division of tensor $B^{\alpha\beta\gamma}$ in 
irreducible parts. First, let us divide tensor $B^{\alpha\beta\gamma}$
in two parts one of which is symmetric and another antisymmetric
in indexes ${\alpha\beta}$
$$B^{\alpha\beta\gamma} =
B^{\left\{\alpha\beta\right\}\gamma}+
B^{\left[\alpha\beta\right]\gamma}~.$$
With the help of invariant tensor $\epsilon_{\alpha\beta\gamma}$ 
the antisymmetric part can be presented in the equivalent form
$$B^{\left[\alpha\beta\right]\gamma}\rightarrow
{B^\gamma}_{\gamma^\prime} = \epsilon_{\alpha\beta{\gamma^\prime}}
B^{\left[\alpha\beta\right]\gamma}~.$$
Like tensor ${Q^\alpha}_\beta$ for mesons tensor ${B^\gamma}_{\gamma^\prime}$
is divided in octet and singlet parts
$${B^\gamma}_{\gamma^\prime} = {B^\gamma_0}_{\gamma^\prime} +
\frac13 {\delta^\gamma}_{\gamma^\prime}\epsilon_{\alpha\beta\delta}
B^{\left[\alpha\beta\right]\delta}~.$$
Coming back to tensor
$$B^{\left[\alpha\beta\right]\gamma} =
\frac12\epsilon^{\alpha\beta{\gamma^\prime}}{B^\gamma}_{\gamma^\prime},$$
we see that the singlet is described by tensor
$$\frac16\epsilon^{\alpha\beta\gamma}B,$$
where
$$B =
\epsilon_{\alpha\beta\gamma}B^{\left[\alpha\beta\right]\gamma}.$$
So, the singlet is presented by totally antisymmetric part of
the initial tensor $B^{\alpha\beta\gamma}$.

The symmetric part  $B^{\left\{\alpha\beta\right\}\gamma}$ can be
further symmetrized or antisymmetrized in indexes $\beta\gamma$
$$ B^{\left\{\alpha\beta\right\}\gamma} =
B^{\left\{\alpha\beta\gamma\right\}} +
B^{\left\{\alpha\left[\beta\right\}\gamma\right]}~.               $$
Symmetrization in indexes $\beta\gamma$ gives totally symmetric
tensor $B^{\left\{\alpha\beta\gamma\right\}}$ ten independent
components of which form the irreducible representation of
baryons- decuplet. Antisymmetrization in indexes $\beta\gamma$ gives
tensor $B^{\left\{\alpha\left[\beta\right\}\gamma\right]}$ that
can be transformed to the equivalent tensor
${B^{\prime\alpha}}_{\alpha^\prime}$ as follows
$$B^{\left\{\alpha\left[\beta\right\}\gamma\right]}\rightarrow
{B^{\prime\alpha}}_{\alpha^\prime} =
\epsilon_{{\alpha^\prime}\beta\gamma}
B^{\left\{\alpha\left[\beta\right\}\gamma\right]}~.$$
Because the trace of tensor ${B^{\prime\alpha}}_{\alpha^\prime}$ 
is equal to zero (${B^{\prime\alpha}}_{\alpha} = \epsilon_{{\alpha}\beta\gamma}
B^{\left\{\alpha\left[\beta\right\}\gamma\right]} = 0$)
it represents a pure octet.

Finally, we have divided the wave function $B^{\alpha\beta\gamma}$ in
symmetric decuplet, antisymmetric singlet and two octets with mixed
symmetry. Let us write the result in the following form
$$3\times3\times3 = \begin{tabular}{cccc}
S&M&M&A\\10~+&$8^\prime~+$&8~+&1\\  & & &  \end{tabular}.$$
If one compares this result with the experimentally observed
multiplets of baryon ground states (decuplet of baryons with
spin 3/2 and octet of baryons with spin 1/2) it can be noted
that the octet of baryons is singly presented and the singlet
baryon is absent at all. As we will see below the reason of
such incompleteness is the Pauli principle for constituent quarks.
Let us write the baryon wave function in the following form
$$\Psi = \begin {array}{c} S~~~~~~~~~~~~~~~~~A~~~~~~~~~~~~~S~~~~~\\
\Psi(x_1,x_2,x_3)\times\Psi(c_1,c_2,c_3)\times B^{\alpha\beta\gamma}
\times\Psi(s_1,s_2,s_3)\\ ~~~~~~~~~~~~~~~~~~~~~\end{array},$$
where the first multiplier is the coordinate wave function that
is assumed to be symmetric for the ground state of the baryon;\\
the second multiplier is the color wave function that is
assumed to be antisymmetric in $SU(3)$ color indexes $c_1,c_2,c_3$
(baryons like all hadrons are assumed to be singlets of
$SU(3)$ - color group; compare the antisymmetry of color wave function
for $SU(3)$ color singlet with the antisymmetry of flavor wave function
for $SU(3)$ flavor singlet);\\
the third multiplier $B^{\alpha\beta\gamma}$ is flavor wave function
the symmetry of which was just described above;\\
the fourth multiplier $\Psi(s_1,s_2,s_3)$ is the spin wave function
the symmetry of which can be described in the same way as the 
symmetry of flavor wave function
$$2\times2\times2 = \begin{tabular}{ccc}
S&M&M\\4~+&$2^\prime~+$&2~\\  & &   \end{tabular}.$$
The antisymmetry of the total wave function (Pauli principle)
will be reached if the product of flavor and spin wave functions
is symmetric. One way to make this product symmetric is
obvious: the decuplet of spin 3/2. Another way, the octet of
spin 1/2, is not so obvious. Let us clarify it by the following
simple argument. If one considers the flavor and spin variables
simultaneously the common index will take 6 values: $u,d,s$ - quarks
with up and down spins. The symmetric in common flavor-spin
indexes tensor (flavor-spin wave function) contains 
$\frac{6\cdot 7\cdot 8}{1\cdot 2\cdot 3} = 56$ independent components.
Forty of them correspond to the decuplet of spin 3/2 ($10\cdot 4$).
There remain 16 components that cannot not to represent the
octet of spin 1/2 ($16 = 8\cdot 2$).\\
So, the Pauli principle selects the possible flavor multiplets
and spins of ground state baryons. Note once more that there is
no place for flavor singlet among the ground state baryons.

\subsection{Two presentations of baryon octet}

As we have seen in the example of mesons the octet of particles
can be described by the traceless matrix $3\times 3$. Let us describe
in such a way the octet of baryons. The problem what particle
should be putted in one or another place in the matrix can be
solved uniquely by the comparison with meson matrix and by the
observation that the third component of isotopic spin 
and the electric charge of compared particles should be equal
($T_3$ and $Q = T_3 + Y/2$ are both the generators of $SU(3)$ - group)
Hence we have the correspondence
$$
\left(\begin{tabular}{ccc}
$\frac{\pi_0}{\sqrt{2}}+\frac{\eta_8}{\sqrt{6}}$&$\pi^+$&$K^+$\\
$\pi^-$ &$ -\frac{\pi_0}{\sqrt{2}}+\frac{\eta_8}{\sqrt{6}}$&$K^0$\\
$K^-$&${\bar K}^0$&$-\frac{2\eta_8}{\sqrt{6}}$ \end{tabular}\right)
\rightarrow
\left(\begin{tabular}{ccc}
$\frac{\Sigma_0}{\sqrt{2}}+\frac{\Lambda}{\sqrt{6}}$&$\Sigma^+$&$p$\\
$\Sigma^-$ &$
-\frac{\Sigma_0}{\sqrt{2}}+\frac{\Lambda}{\sqrt{6}}$&$n$\\
$\Xi^-$&${\Xi}^0$&$-\frac{2\Lambda}{\sqrt{6}}$
\end{tabular}\right)$$
This description of baryon octet will be used in the following lecture
for the derivation of mass formulae for baryon octet.

Another presentation of baryon octet refers directly to the quark
structure of baryons. Let us construct the  proton state, for example.
The proton consists of two $u$ -quarks and one $d$ - quark i.e.
its flavor composition is $uud$. According to Pauli principle
the total spin of two $u$ -quarks should be equal to 1 (see the
symmetry of flavor-spin wave function stated above). Let us sum
the unit spin  of two $u$ -quarks with 1/2 spin of $d$ - quark
to obtain 1/2 spin of proton
$$
{\dot p} = \sqrt{\frac23} {\dot u}{\dot u}\d{d}-
\sqrt{\frac13}\sqrt{\frac12}({\dot u}\d{u} + \d{u}{\dot u}){\dot d}
$$
(here the point above (below) the letter means spin up (down) for
quark) and symmetrize the obtained state by permutations of
$d$ - quark
$$
{\dot p} = \sqrt{\frac{1}{18}}(2({\dot u}{\dot u}\d{d}+
{\dot u}\d{d}{\dot u}+\d{d}{\dot u}{\dot u})-({\dot u}\d{u}{\dot d}+
{\dot u}{\dot d}\d{u}+{\dot d}{\dot u}\d{u}+\d{u}{\dot u}{\dot d}+
\d{u}{\dot d}{\dot u}+{\dot d}\d{u}{\dot u}))~~~~.$$
The obtained description of baryon octet will be used in the
lecture on magnetic moments of baryons.

\subsection{Problem 5}

Classify the baryons with one $c$ - quark ($cqq$) in  possible
$SU(3)$ - multiplets and  spin.
\newpage
\section{Lecture 6. Mass formulae. Mixing}

Last lecture we were classifying the $SU(3)$ - multiplets of
ground states of mesons $q{\bar q}$ and baryons $qqq$. These
were the octets and singlets of pseudoscalar and vector mesons
and the octet and decuplet of baryons. If $SU(3)$ - symmetry 
were exact symmetry the masses of particles in each multiplet
would be equal to each other. In the real world the 
$SU(3)$ - symmetry is an approximate symmetry  and
the masses of particles in the multiplets are essentially different:
$$ {P^\alpha}_\beta (\begin{array}{c}
\pi~~~~~~K~~~~~~\eta~~~~~~\eta^\prime\\
140~~~490~~~550~~~960\\~~~\end{array})$$
$$ {V^\alpha}_\beta (\begin{array}{c}
\rho~~~~~~K^*~~~~~~\omega~~~~~~\phi\\
770~~~890~~~780~~~1020\\~~~\end{array})$$
$$ {B^\alpha}_\beta (\begin{array}{c}
N~~~~~~\Lambda~~~~~~\Sigma~~~~~~\Xi\\
940~~~1116~~~1190~~~1320\\~~~\end{array})$$
$$ D^{\alpha\beta\gamma} (\begin{array}{c}
\Delta~~~~~~\Sigma^*~~~~~~\Xi^*~~~~~~\Omega\\
1230~~~1380~~~1530~~~1670\\~~~\end{array})$$

The violation of $SU(3)$ - symmetry initiates on the
quark level and is connected to the mass difference
of nonstrange $u,d$ - quarks ($4,7 MeV$) and strange $s$ - quark
($150 MeV$). The consequences of just this mechanism of
symmetry violation are the mass formulae that can be
deduced by considerations of symmetry properties of quark
Lagrangian only.\\
On the quark level the  Lagrangian violating the $SU(3)$ - symmetry
is mass Lagrangian of quarks
$$L^q_m = m_q({\bar u}u + {\bar d}d) + m_s{\bar s}s = m_q({\bar u}u
 + {\bar d}d + {\bar s}s) + (m_s-m_q){\bar s}s.$$
Here we have neglected the violation of isotopic symmetry
connected to the mass difference
of nonstrange $u,d$ - quarks ($4,7 MeV$). Its account
(see the problem in the end of this lecture) without
an account of electromagnetic interaction violating
both $SU(3)$ and isotopic symmetries would be unjustified.

The mass Lagrangian of quarks is presented by the sum of
two terms: $SU(3)$ symmetrical term (singlet under $SU(3)$ - group
transformation) and $SU(3)$ violating term (transforming like
(3,3) component of the tensor under $SU(3)$ - group
transformation)
$$L^{q0}_m =  m_q({\bar u}u + {\bar d}d + {\bar s}s)$$
$$L^{q3}_{m3} = (m_s-m_q){\bar s}s.$$
It is reasonable to expect that on the hadronic level the
mass Lagrangians of hadrons will also contain two terms
obeying the same transformation properties under $SU(3)$ - group
transformations. 
 
\subsection{Octet of baryons}

First, let us consider the octet of baryons with spin 1/2
$${B^\alpha}_\beta = \left(\begin{tabular}{ccc}
$\frac{\Sigma_0}{\sqrt{2}}+\frac{\Lambda}{\sqrt{6}}$&$\Sigma^+$&$p$\\
 $\Sigma^-$ &$
-\frac{\Sigma_0}{\sqrt{2}}+\frac{\Lambda}{\sqrt{6}}$&$n$\\
$\Xi^-$&${\Xi}^0$&$-\frac{2\Lambda}{\sqrt{6}}$
\end{tabular}\right)~.$$
The singlet under $SU(3)$ - group
transformations mass Lagrangian of baryon octet is unique -
the masses of all baryons of the octet ($m^0_B$) are equal in
this Lagrangian
$$L^{B0}_m = m^0_B {{\bar B}^\alpha}~ _\beta {B^\beta}_\alpha =
m^0_B({\bar p}p + {\bar n}n + \cdot\cdot\cdot + {\bar \Lambda}
\Lambda)~.$$
The $SU(3)$ violating part of mass  Lagrangian of baryon octet,
(3,3) component of the tensor, can be presented in two forms,
hence, there are two independent mass  parameters
($m^8_B$ and ${m^8_B}^\prime$)
$$\begin{array}{c} L^{B3}_{m3} = m^8_B {{\bar B}^3}~_\beta {B^\beta}_3
+ {m^8_B}^\prime {{\bar B}^\alpha}~_3 {B^3}_\alpha =\\
m^8_B({\bar p}p + {\bar n}n + \frac23 {\bar \Lambda}\Lambda) +
 {m^8_B}^\prime ({\bar \Xi}^-\Xi^- + {\bar \Xi}^0\Xi^0 +
\frac23 {\bar \Lambda}\Lambda)\end{array}$$
The masses of baryons are contributed by both $L^{B0}_m $ and
$L^{B3}_{m3}$
$$m_p = m_n = m^0_B + m^8_B$$
$$m_{\Xi^-} = m_{\Xi^0} = m^0_B + {m^8_B}^\prime$$
$$m_{\Sigma^+} = m_{\Sigma^0} = m_{\Sigma^-} = m^0_B$$
$$m_\Lambda = m^0_B + \frac23(m^8_B + {m^8_B}^\prime)~.$$
So, four baryon masses (the violation of isotopic symmetry is
not considered) are expressed through three parameters. Thus,
there exits the constraint
$$3m_\Lambda + m_\Sigma = 2(m_N + m_\Xi),$$
the so called Gell-Mann - Okubo mass formula.
The left handed part of this formula is equal to
$$3\cdot 1116 + 1193 = 3348 +1193 = 4541,$$
whereas the right handed part of this formula is equal to
$$2\cdot (939 + 1318) = 2\cdot 2257 = 4517.$$
If we have used the Gell-Mann - Okubo mass formula for
the prediction of the mass of $\Lambda$ - hyperon we
would obtain the result $m_\Lambda = 1107 MeV$ that is only
$9 MeV$ deviated from the experimental value. The difference
is comparable with electromagnetic/isotopic mass differences.
Thus we can conclude that the accuracy of Gell-Mann - Okubo mass formula
is very large.

\subsection{Decuplet of baryons}

Let us now consider the decuplet of baryons $D^{\alpha\beta\gamma}$.
The invariant part of mass Lagrangian for decuplet particles
can be written as follows
$$\begin{array}{c}L^{D0}_m = m^0_D {\bar
D}_{\alpha\beta\gamma}D^{\alpha\beta\gamma} = m^0_D \{(111) + (222) +
(333)\\ + 3[(112) + (113) + (221) + (223) + (331) + (332)] +
 6(123)\},\end{array}$$ 
where the notation  $(\alpha\beta\gamma)$ means 
${\bar D}_{\alpha\beta\gamma}D^{\alpha\beta\gamma}$ with no
summation over indexes. Taking into account the equality
of the masses of decuplet particles in this invariant part of mass Lagrangian
we obtain the following correspondence between the components
of tensor $D^{\alpha\beta\gamma}$ and the fields of the decuplet particles
$$\begin{array}{l} \Delta^{++}\leftrightarrow D^{111}\\
\frac{\Delta^+}{\sqrt{3}}\leftrightarrow D^{112}\\
\frac{\Delta^0}{\sqrt{3}}\leftrightarrow D^{221}\\
\Delta^{-}\leftrightarrow D^{222}\end{array}
\begin{array}{c}\frac{\Sigma^{+*}}{\sqrt{3}}\leftrightarrow D^{113}\\
\frac{\Sigma^{0*}}{\sqrt{6}}\leftrightarrow D^{123}\\
\frac{\Sigma^{-*}}{\sqrt{3}}\leftrightarrow D^{223}\end{array}
\begin{array}{c}
\frac{\Xi^{0*}}{\sqrt{3}}\leftrightarrow D^{331}\\
\frac{\Xi^{-*}}{\sqrt{3}}\leftrightarrow D^{332}\\\end{array}
\Omega^-\leftrightarrow D^{333}~.$$
$SU(3)$ violating part of mass Lagrangian for decuplet particles
is described by the unique structure
$$L^{D3}_{m3} = m^8_D {\bar
D}_{3\alpha\beta}D^{3\alpha\beta} = m^8_D \{(311) + (322) + (333) +
2[(312) + (313) + (323)]\}~.$$
Taking into account the above stated correspondence between the components
of tensor $D^{\alpha\beta\gamma}$ and the fields of the decuplet particles
we get for the masses of decuplet particles the following results
$$m_\Delta = m^0_D~~~~~~(1230)$$
$$m_{\Sigma^*} = m^0_D + m^8_D/3~~~~~~(1380)$$
$$m_{\Xi^*} = m^0_D + 2m^8_D/3~~~~~~(1520)$$
$$m_{\Omega} = m^0_D + m^8_D~~~~~~(1670)~.$$
The masses of particles are equidistant in a good agreement
with the experiment.

\subsection{Mesons. Mixing}

Turning now to mesons let us note two peculiarities of
this case. First, the mass Lagrangian of mesons contains as
parameters the masses of mesons squared in contrast with
baryon case where the mass Lagrangian was linear in mass parameters.
Therefore the mass formulae for mesons will relate the masses
of mesons squared. So, the Gell-Mann - Okubo formula transforms
in the case of pseudoscalar and vector mesons to the following
formulae
$$3m^2_{\eta_8} + m^2_{\pi} = 4m^2_K$$
$$3m^2_{\omega_8} + m^2_{\rho} = 4m^2_{K^*}~.$$
Here $\eta_8$ and $\omega_8$ are isotopic singlet components of
unitary octets
$${P^3}_3 = -\frac{2\eta_8}{\sqrt{6}}~~~~~~~~
{V^3}_3 = -\frac{2\omega_8}{\sqrt{6}}.$$ 
This particles are not observed experimentally like
are not observed also isotopic and unitary singlets $\eta_0$ and $\omega_0$.
The reason of their inobservability (the second peculiarity of
mesons in comparison with baryons) is that $SU(3)$ singlet and octet
mesons can mix due to $SU(3)$ - symmetry violation. The mixing term
in the mass Lagrangian of mesons has the transformation property
of (3,3)-component of tensor i.e. is allowed by symmetry reasons
$$L^{mix}_m = m^2_{mix} {M^3}_3M_0,$$
where ${M^\alpha}_\beta$ is the tensor of the meson octet and
$M_0$ is the singlet meson. The result of mixing of mesons
$\eta_8$ and $\eta_0$ ($\omega_8$ and $\omega_0$) are the mesons
with definite mass $\eta$ and $\eta^\prime$ ($\phi$ and $\omega$).
Let us describe the mixing process.

Let us picture the axis where the square of particle masses
will be indicated. First, consider the pseudoscalar mesons.
Due to the mass formula we know the mass squared of $\eta_8$ - meson:
 $m^2_{\eta_8} = (4m^2_K - m^2_{\pi})/3 = (566 MeV)^2$. Experimentally
 the masses squared of $\eta$ and $\eta^\prime$ - mesons are known:
 $m^2_\eta = (549 MeV)^2$ and $m^2_{\eta^\prime} =
 (958 MeV)^2$
 $$------|------|-----------|------|------>
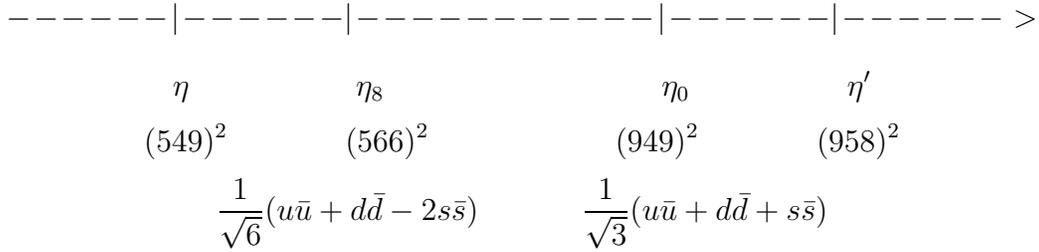
$$
$$~~~~~~~~~~~~~~~~~~\eta~~~~~~~~~~~~~~~~~~\eta_8~~~~~~~~~~~~~~~~~~~~~~~~~~~~~~
\eta_0~~~~~~~~~~~~~~~~~\eta^\prime~~~~~~~~~~~~~~~~~~$$
$$~~~~~~~~~~~~~~(549)^2~~~~~~~~~~~~(566)^2~~~~~~~~~~~~~~~~~~~
(949)^2~~~~~~~~~~~~(958)^2~~~~~~~~~~~~~~$$
$$~~~~~~~~~~{\small \frac{1}{\sqrt{6}}(u{\bar u} + d{\bar d} -
2s{\bar s})} ~~~~~~~~~~{\small \frac{1}{\sqrt{3}}(u{\bar u} + d{\bar
d} + s{\bar s})}~~~~~~~~~~$$
\vspace{1 cm}
The mass squared of $\eta_0$ на meson can be found by
the formula of two level mixing theory:
$$m^2_{\eta^\prime} - m^2_{\eta_0} =
m^2_{\eta_8} - m^2_\eta$$ ( two levels are repelling in opposite
directions and on equal distances). 
We see that the nondiagonal $\eta_8$ - meson is lighter than the
nondiagonal $\eta_0$ - meson although $\eta_8$ - meson contains
more heavy strange quarks than $\eta_0$ - meson and this is
a surprise that will be discussed in the next lecture. The mixing
of pseudoscalar mesons is described by the mixing angle $\theta_P$
$$\eta = cos\theta_P \eta_8 + sin\theta_P \eta_0$$
$$\eta^\prime = -sin\theta_P \eta_8 + cos\theta_P \eta_0,$$
that can be found one inverts the mixing formulae
$$\eta_8 = cos\theta_P \eta - sin\theta_P \eta^\prime$$
$$\eta_0 = sin\theta_P \eta + cos\theta_P \eta^\prime$$
and one calculates the matrix element of mass squared operator,say,
over the state of $\eta_8$ - meson
$$m^2_{\eta_8} = cos^2\theta_P m^2_\eta + sin^2\theta_P
m^2_{\eta^\prime}. $$ 
Then we obtain
$$sin^2\theta_P = \frac{m^2_{\eta_8} - m^2_\eta}{m^2_{\eta^\prime} -
m^2_\eta}~~~~~~~\rightarrow |\theta_P| \approx 10^o~.$$
The mass formulae prediction of  the mixing angle $\theta_P$
will be discussed in the following lecture where the annihilation
of pseudoscalar mesons into two photons will be considered.

Let us turn now to vector mesons. We know the mass squared of
$\omega_8$ - meson: $m^2_{\omega_8} = (4m^2_{K^*} - m^2_{\rho})/3 = (929
MeV)^2$. The masses squared of diagonal $\omega$ and $\phi$ - mesons
are known experimentally:  $m^2_\omega = (780 MeV)^2$ и $m^2_{\phi} =
 (1020 MeV)^2$
 $$------|------|-----------|------|------>$$
$$~~~~~~~~~~~~~~~~~~\omega~~~~~~~~~~~~~~~~~~\omega_0~~~~~~~~~~~~~~~~~~~~~~~~~~~~~~
\omega_8~~~~~~~~~~~~~~~~~\phi~~~~~~~~~~~~~~~~~~$$
$$~~~~~~~~~~~~~~(780)^2~~~~~~~~~~~~(900)^2~~~~~~~~~~~~~~~~~~~
(929)^2~~~~~~~~~~~~(1020)^2~~~~~~~~~~~~~~$$
$$~~~~~~~~~~{\small \frac{1}{\sqrt{3}}(u{\bar u} + d{\bar d} +
 s{\bar s})} ~~~~~~~~~~{\small \frac{1}{\sqrt{6}}(u{\bar u} + d{\bar
d} - 2s{\bar s})}~~~~~~~~~~$$
\vspace{1 cm}
The mass of $\omega_0$ - meson obtained by the use of
mixing formulae is less than the mass of $\omega_8$ - meson
and this is not a surprise because $\omega_0$ - meson contains
less strange quarks than  $\omega_8$ - meson.

The mixing of vector mesons is described by the mixing angle $\theta_V$
$$\omega = cos\theta_V \omega_0 + sin\theta_V \omega_8$$
$$\phi = -sin\theta_V \omega_0 + cos\theta_V \omega_8.$$
According to mass formulae this angle is equal to
$$sin^2\theta_V = \frac{m^2_{\omega_0} -
m^2_\omega}{m^2_{\phi} - m^2_\omega}~~~~~~~\rightarrow
~~~~~~~|\theta_V| \approx 40^o.$$
The angle $\theta_V$ is close to the ideal mixing angle
$$cos {\theta_V}_{ideal} = \sqrt{\frac23}~~~~~~~~
sin {\theta_V}_{ideal} = \sqrt{\frac13}~~~~~~~~{\theta_V}_{ideal}\approx
 35^o,$$
 for which $\omega$ -  meson contains only nonstrange quarks
 and $\phi$ - meson contains only strange quarks.
In more detail the mixing of mesons will be discussed in the
following lecture.

\subsection{Problem 6}

Find the mass formulae for the baryon octet taking into account
the violation  not only of $SU(3)$ - symmetry but also of
$SU(2)$ - symmetry.
\newpage
\section{Lecture 7. Mixing of pseudoscalar and vector mesons}
\subsection{ $SU(3)$ - symmetry limit for the masses of
pseudoscalar and vector mesons}

In the previous lecture we have considered the effects of
$SU(3)$ - symmetry violation in the masses of mesons and
baryons. The result was the Gell-Mann - Okubo mass formulae.
In the case of vector mesons the mass formula has predicted
the mass of nondiagonal $\omega_8$ - meson to be $m_{\omega_8} = 929 MeV$.
The mixing theory then has predicted
the mass of nondiagonal $\omega_0$ - meson to be $m_{\omega_0} = 900 MeV$:
for this mass of nondiagonal  $\omega_0$ - meson  the mixing of
$\omega_8$ and $\omega_0$ - mesons produces the diagonal $\phi$ and
$\omega$- mesons with observed masses. The result that $\omega_8$ - meson 
is more heavy than $\omega_0$ - meson was not an unexpected one  because
$\omega_8$ - meson contains more strange quarks than $\omega_0$ - meson.
Then it is natural to expect that the decreasing of strange quark mass
to nonstrange quark mass would result in the degeneracy of 
$\omega_8$ and $\omega_0$ - mesons ($m_{\omega_8} \rightarrow m_{\omega_0}$)
and that in the $SU(3)$ - symmetry limit we would have the nonet
(octet plus singlet) of degenerate vector mesons.\\
On the contrary in the case of pseudoscalar mesons
$m_{\eta_8} << m_{\eta_0}$ although $\eta_8$ - meson contains
more strange quarks than $\eta_0$ - meson. 
Then it is natural to expect that the decreasing of strange quark mass
to nonstrange quark mass would result in the increase of 
$m_{\eta_0} - m_{\eta_8}$ mass difference and that in the 
$SU(3)$ - symmetry limit we would have the octet of pseudoscalar
mesons with masses $\sim m_\pi$  pseudoscalar singlet with mass
$m_{\eta_0} >> m_{\pi}$. For this reason it is widely accepted
to think that some part of $\eta_0$ - meson (and hence of
$\eta^\prime$ - meson) is presented by the gluonic component i.e.
 $$\eta_0 = \alpha
\frac{1}{\sqrt{3}}(u{\bar u} + d{\bar d} + s{\bar s}) + \beta
G{\tilde G},$$
where the symbol $G{\tilde G}$ denotes the pseudoscalar glueball 
with quantum numbers $J^{PC} = 0^{-+}$.

\subsection{Mixing of pseudoscalar mesons}

The information on the pseudoscalar mesons mixing angle $\theta_P$
defined by the formulae
$$\eta = cos\theta_P \eta_8 + sin\theta_P \eta_0$$
$$\eta^\prime = -sin\theta_P \eta_8 + cos\theta_P \eta_0,$$
can be obtained not only from mass formulae where $|\theta_P| \approx
10^o$ but also from the decays of pseudoscalar mesons .
There are three pseudoscalar mesons decaying to two photons -
$\pi_0,\eta,\eta^\prime $. The amplitude of the transition of
$q{\bar q}$ pair to two photons is proportional to the 
electric charge of $q$ - quark squared $e^2_q \equiv g_{q{\bar q}}$ .
In the limit of exact $SU(3)$ - symmetry the amplitudes of
$\pi_0,\eta_8,\eta_0 \rightarrow 2\gamma$ transitions would
differ only by the values of $g_{\pi_0},g_{\eta_8},g_{\eta_0}$
constants
$$\begin{array}{c}
g_{\pi_0} = \frac{1}{\sqrt{2}}(e^2_u - e^2_d) =
\frac{1}{\sqrt{2}}(\frac49 - \frac19) = \frac{1}{\sqrt{2}~~3}\\
g_{\eta_8} = \frac{1}{\sqrt{6}}(e^2_u + e^2_d - 2e^2_s) =
\frac{1}{\sqrt{6}}(\frac49 + \frac19 - 2\frac19) =
\frac{1}{\sqrt{6}~~3}\\
g_{\eta_0} = \frac{1}{\sqrt{3}}(e^2_u + e^2_d + e^2_s) =
\frac{1}{\sqrt{3}}(\frac49 + \frac19 + \frac19) =
\frac{2}{\sqrt{3}~~3}\end{array}.$$
The violation of $SU(3)$ - symmetry contributes at least two
modifications. First, the decaying particles  $\eta$ and $\eta^\prime$
are the mixtures of $\eta_8$ and $\eta_0$ and
$$\begin{array}{c}g_\eta = cos\theta_P g_{\eta_8} + sin\theta_P
g_{\eta_0} = \frac13(\frac{1}{\sqrt{6}} cos\theta_P +
\frac{2}{\sqrt{3}} sin\theta_P)\\g_{\eta^\prime} = -sin\theta_P
g_{\eta_8} + cos\theta_P g_{\eta_0} = \frac13(-\frac{1}{\sqrt{6}}
sin\theta_P + \frac{2}{\sqrt{3}} cos\theta_P)\end{array}.$$
Second, the amplitudes and the widths of the considered decays are
different
because the decaying particles have different masses, different energies
of photons. Let us take into account the dependence on the mass
of decaying pseudoscalar meson phenomenologically
$$\begin{array}{c} A(P\rightarrow 2\gamma) \sim g_P P F_{\mu\nu}
{\tilde F}_{\mu\nu} \sim g_P m^2_P\\
\Gamma(P\rightarrow 2\gamma) \sim |A(P\rightarrow
2\gamma)|^2\frac{1}{2m_P} \sim g^2_P m^3_P\end{array}.$$
Here (${\tilde F}_{\mu\nu}) F_{\mu\nu}$ (dual) tensor of
electromagnetic field. So, we have the ratios
$$\frac{\Gamma_\eta}{\Gamma_{\pi_0}} = \frac{g^2_\eta}{g^2_{\pi_0}}
\frac{m^3_\eta}{m^3_{\pi_0}} = (cos\theta_P + 2\sqrt{2}sin\theta_P)^2
\frac13\frac{m^3_\eta}{m^3_{\pi_0}}~,$$
that predicts $\theta_P \approx 11.8^o$ and
$$\frac{\Gamma_{\eta^\prime}}{\Gamma_{\pi_0}} =
\frac{g^2_{\eta^\prime}}{g^2_{\pi_0}}
\frac{m^3_{\eta^\prime}}{m^3_{\pi_0}} = (-sin\theta_P +
2\sqrt{2}cos\theta_P)^2 \frac13\frac{m^3_{\eta^\prime}}{m^3_{\pi_0}}~,$$
that predicts $\Gamma^{th}_{\eta^\prime} = 783 \Gamma_{\pi_0} = 6.07 KeV (
\Gamma^{exp}_{\eta^\prime} = 4.54 KeV )$. If the second ratio was
used for the determination of angle $\Theta_P$ the result would be
$\theta_P \approx 22.8^o$ in disagreement with the result
 $|\theta_P| \approx 10^o$ of mass formulae. The possible reason
 of this disagreement can be the presence of gluonic component
 in $\eta^\prime$ - meson discussed above. Let us assume that in
 the mixture
 $$\eta_0 = \alpha
\frac{1}{\sqrt{3}}(u{\bar u} + d{\bar d} + s{\bar s}) + \beta
G{\tilde G}$$
the  gluonic component does not annihilates to two photons. Then
the width $\Gamma^{th}_{\eta^\prime}$  will be equal to
$$\Gamma^{th}_{\eta^\prime} = \alpha^2 6.07 KeV$$
and $\alpha^2 \approx 75 \%$ to explain the experimental width
$\Gamma^{exp}_{\eta^\prime} = 4.54 KeV$. Thus we have the second
argument in favor of gluonic component in $\eta^\prime$ - meson.
The nature of this gluonic component is a separate interesting
problem that we will not discuss here.

\subsection{Mixing of vector mesons}

Let us now consider the mixing of vector mesons. According to
mass formulae the mixing
$$\omega = cos\theta_V \omega_0 + sin\theta_V \omega_8$$
$$\phi = -sin\theta_V \omega_0 + cos\theta_V \omega_8$$
with $|\theta_V| \approx 40^o$ is close to the ideal mixing
with ${\theta_V}_{ideal}\approx 35^o$ for which $\omega$ - meson
does not contain strange quarks and $\phi$ - meson
does not contain nonstrange quarks. Like the case of pseudoscalar
mesons the mixing of vector mesons can be studied in electromagnetic
decays of vector mesons - annihilation to electron-positron pair.
The amplitude of $q{\bar q}$ - pair annihilation to electron-positron pair
is proportional to the quark charge $e_q \equiv h_{q{\bar q}}$ . In the
case of exact $SU(3)$ - symmetry we would have diagonal
${\rho_0},{\omega_8},{\omega_0}$ - mesons and their couplings
$$\begin{array}{c} h_{\rho_0} = \frac{1}{\sqrt{2}}(\frac23 -
(-\frac13)) = \frac{1}{\sqrt{2}}\\
h_{\omega_8} = \frac{1}{\sqrt{6}}(\frac23 + (-\frac13) - 2(-\frac13))
= \frac{1}{\sqrt{6}}\\
h_{\omega_0} = \frac{1}{\sqrt{3}}(\frac23 + (-\frac13) + (-\frac13))
= 0.\end{array}$$
Like the case of pseudoscalar
mesons the violation of $SU(3)$ - symmetry also contributes two
modifications here. First, we have the  mixing of vector mesons
and 
$$\begin{array}{c} h_{\omega} = cos\theta_V h_{\omega_0} +
sin\theta_V h_{\omega_8} = sin\theta_V\frac{1}{\sqrt{6}}\\ h_{\phi} =
-sin\theta_V h_{\omega_0} + cos\theta_V h_{\omega_8} =
cos\theta_V\frac{1}{\sqrt{6}}.\end{array}$$
Second, the decay width $\Gamma(V \rightarrow e^+ e^-) $ depend on
the mass of vector meson. This dependence can be found by
the following arguments. The widths of the annihilation of vector
meson is proportional to the probability for quark and antiquark
to meet - the square of wave function of quark and antiquark
at origin. Making the desired dimensionality for the width
by the use of the proper power of the vector meson mass we get
$$\Gamma(V \rightarrow e^+ e^-) \sim \frac{|\psi_V(0)|^2}{m^2_V}
h^2_V.$$
Finally we have the ratio
$$\frac{\Gamma_\omega}{\Gamma_\rho} = \frac{sin^2\theta_V}{3}
\frac{m^2_\rho}{m^2_\omega},$$
from which $\theta_V \approx 32^o$ close to ${\theta_V}_{ideal}\approx 35^o$
can be obtained and the ratio
$$\frac{\Gamma_\phi}{\Gamma_\rho} = \frac{cos^2\theta_V}{3}
\frac{m^2_\rho}{m^2_\phi},$$
from which one follows that $\Gamma^{th}_\phi \approx 1 KeV$. The $30\%$
difference with the experimental width $\Gamma^{exp}_\phi \approx 1.3 KeV$
can be explained by larger value of the wave function at the origin
in the case of strange (heavy) quarks compared to the case of
nonstrange (light) quarks.

\subsection{Problem 7}

Make all the numerical calculations omitted in the lecture.

\newpage
\section{Lecture 8. Magnetic moments of baryons}
\subsection{$SU(3)$ symmetric limit for magnetic moments of baryons }

$SU(3)$ - symmetry of strong interaction predicts the relations
for various physical characteristics of hadrons. One of the most
impressive example is the magnetic moments of baryons . Like the
case of mass formulae let us start from the quark level - from
electromagnetic current of quarks. In the limit of $SU(3)$ - symmetry
$u,d,s$ - quarks differ only by the values of  their electric charges.
The electromagnetic current of quarks has the following $SU(3)$ - 
structure
$$\begin{array}{l}j_\mu = \frac23{\bar u}\gamma_\mu u - \frac13{\bar
d}\gamma_\mu d - \frac13{\bar s}\gamma_\mu s =\\
= {\bar u}\gamma_\mu u - \frac13({\bar u}\gamma_\mu u + {\bar
d}\gamma_\mu d + {\bar s}\gamma_\mu s) \end{array}$$
or
$$j_\mu = J^1_{\mu 1} - \frac13(J^\alpha_{\mu \alpha}).$$
Considering the hadrons we will construct the  electromagnetic current of
hadrons that possesses the same transformation properties with
respect to $SU(3)$ - transformations as electromagnetic current of quarks.
In particular, for the octet of baryons that is described
by the matrix
$${B^\alpha}_\beta = \left(\begin{tabular}{ccc}
$\frac{\Sigma_0}{\sqrt{2}}+\frac{\Lambda}{\sqrt{6}}$&$\Sigma^+$&$p$\\
 $\Sigma^-$ &$
-\frac{\Sigma_0}{\sqrt{2}}+\frac{\Lambda}{\sqrt{6}}$&$n$\\
$\Xi^-$&${\Xi}^0$&$-\frac{2\Lambda}{\sqrt{6}}$
\end{tabular}\right),$$
the electromagnetic current can be presented by two independent
flavor structures
$$\begin{array}{r}
<j_\mu >_B = {{\bar B}^\alpha}~ _1 \Gamma_\mu {B^1}_\alpha -
\frac13 {{\bar B}^\alpha}~ _\beta \Gamma_\mu {B^\beta}_\alpha +\\
+{{\bar B}^1}~ _\beta \Gamma^\prime_\mu {B^\beta}_1 - \frac13
{{\bar B}^\alpha}~ _\beta \Gamma^\prime_\mu {B^\beta}_\alpha
,\end{array}~,$$
where the quantities $ \Gamma_\mu$ and $\Gamma^\prime_\mu$
are the combinations of electric and magnetic formfactors
$$\begin{array}{c}
\Gamma_\mu = \gamma_\mu f_1(q^2) + \sigma_{\mu\nu} q^\nu f_2(q^2)\\
\Gamma^\prime_\mu = \gamma_\mu f^\prime_1(q^2) + \sigma_{\mu\nu}
q^\nu f^\prime_2(q^2).\end{array}~.$$
Here $q$ is the four-momentum  transfered to the baryon. Electric 
and magnetic formfactors of baryons are the definite combinations
of two electric ($f_1(q^2)$ and $f^\prime_1(q^2)$ ) and
two magnetic ($f_2(q^2)$ and $f^\prime_2(q^2)$ ) formfactors.
In  accordance to this the electric charges and magnetic moments
of baryons are the definite combinations of constants
$Q$,$Q^\prime$ and  $\mu$,$\mu^\prime$ defined as follows
$$\begin{array}{l} Q = f_1(0)~~~~~~~~~~~~Q^\prime = f^\prime_1(0)\\
\mu_A = f_2(0)~~~~~~~~~~~~\mu^\prime_A = f^\prime_2(0)\\
\mu = \frac{Q}{2m} + \mu_A~~~~~~\mu^\prime = \frac{Q^\prime}{2m} +
\mu^\prime_A ,\end{array}~.$$ 
So, magnetic moments of baryons being the combinations of two
independent parameters $\mu$,$\mu^\prime$ can be expressed
through two independent baryon magnetic moments, say, the
magnetic moments of proton and neutron. Thus we get the
following table

\begin{tabular}{cccc}
$\mu_p$~~~~~~~~~&$\mu - \frac13(\mu + \mu^\prime)$~~~~~~~~~ &
input &~~~~~~~~~2.79\\
$\mu_n$~~~~~~~~~&$ - \frac13(\mu + \mu^\prime)$~~~~~~~~~ &
input &~~~~~~~~~-1.91\\
$\mu_{\Sigma^+}$~~~~~~~~~&$\mu - \frac13(\mu +
\mu^\prime) = \mu_p$~~~~~~~~~ & 2.79 &~~~~~~~~~2.42\\
$\mu_{\Sigma^0}$~~~~~~~~~&$(\frac12 - \frac13)(\mu +
\mu^\prime) = -\frac12 \mu_n$~~~~~~~~~ & 0.96 &~~~~~~~~~---\\
$\mu_{\Sigma^-}$~~~~~~~~~&$\mu^\prime - \frac13(\mu +
\mu^\prime) = -(\mu_p + \mu_n)$~~~~~~~~~ & -0.88
&~~~~~~~~~-1.16\\
$\mu_{\Xi^-}$~~~~~~~~~&$\mu^\prime - \frac13(\mu +
\mu^\prime) = -(\mu_p + \mu_n)$~~~~~~~~~ & -0.88
&~~~~~~~~~-0.68\\
$\mu_{\Xi^0}$~~~~~~~~~&$ - \frac13(\mu + \mu^\prime) =
\mu_n$~~~~~~~~~ & -1.91 &~~~~~~~~~-1.25\\
$\mu_{\Lambda}$~~~~~~~~~&$(\frac16 - \frac13)(\mu + \mu^\prime) =
\frac12 \mu_n$~~~~~~~~~ &-0.96 &~~~~~~~~~-0.61\\
$\mu_{\Sigma\Lambda}$~~~~~~~~~&$(\frac{1}{\sqrt{12}}(\mu +
\mu^\prime) = -\frac{3}{\sqrt{12}} \mu_n$~~~~~~~~~ &1.65
&~~~~~~~~~1.61
\end{tabular}

(all numerical values here are expressed in units $\frac{1}{2m}$
where $m$ is the nucleon mass)

\vspace{1cm}

The difference among the $SU(3)$ - predictions for magnetic
moments of baryons (left numbers) and experimental data
(right numbers) are rather large. It means that $SU(3)$ - symmetry
in magnetic
moments of baryons is noticeably violated . It is quite educative
to look for the relations among the  magnetic
moments of baryons in another way that will let both
to take into account the effects of $SU(3)$ - symmetry violation
and relate the  magnetic moments of proton and neutron (they were
not related in the above considerations). The assumption of
additivity of quark magnetic moments in the magnetic moment of baryon
will be the key in this another way.

\subsection{ Magnetic moment of baryon in additive quark model}

The magnetic moment ($\mu$) of spin 1/2 particle is defined by the
relation
$${\vf \mu} = \mu {\vf \sigma},$$
where ${\vf \sigma}$ are Pauli matrices acting on spin variables
of particle. Assuming that baryons consist of three quarks and that
the quark magnetic moments are simply added in the magnetic moment of baryon
we get
$${\vf \mu} = \Sigma_i \mu_i {\vf \sigma}_i~~~~~~~~~~~~(i = 1,2,3)$$
The matrices ${\vf \sigma}_i$ act on spin variables of constituent quarks.
The baryon magnetic moment ($\mu$) is, by definition, the matrix
element of operator $\mu_z$ over the baryon state with spin projection
on z-axis equal to 1/2
$$\mu = <s_z =1/2| \mu_z | s_z = 1/2>.$$
For the calculation of  magnetic moment of baryons we will use
the explicit constructions of baryon states found in lecture 5.
So, for the proton we have
$$
{\dot p} = \sqrt{\frac23} {\dot u}{\dot u}\d{d}-
\sqrt{\frac13}\sqrt{\frac12}({\dot u}\d{u} + \d{u}{\dot u}){\dot
d}~~,  $$
where the exact simmetrization with respect to the third quark
is omitted (this simmetrization does not change the result). Calculating
the corresponding matrix
element of operator $\mu_z$ we get the proton  magnetic moment 
$$\mu_p = \frac23(\mu_u + \mu_u - \mu_d) + \frac26(\mu_u - \mu_u +
\mu_d) = \frac13(4\mu_u - \mu_d).$$
The magnetic moments of six other components of baryon octet follow
immediately from the comparison with the result for
the proton  magnetic moment
$$\mu_n = \frac13(4\mu_d - \mu_u) ~~~~~~~~~~\mu_{\Sigma^+} =
\frac13(4\mu_u - \mu_s)$$
$$\mu_{\Sigma^-} =
\frac13(4\mu_d - \mu_s)~~~~~~~~~~\mu_{\Xi^-} =
\frac13(4\mu_s - \mu_d)$$
$$\mu_{\Xi^0} = \frac13(4\mu_s - \mu_u)~~~~~~~~~~\mu_{\Sigma^0} =
\frac13(2(\mu_u + \mu_d) - \mu_s)~.$$
The separate consideration is needed for $\Lambda$ - hyperon. Its
wave function (without simmetrization with respect to quark interchanges
that does not change the result) has the form
$${\dot \Lambda} = \frac{1}{\sqrt{2}}({\dot u}\d{d} - \d{u}{\dot d})
{\dot s}~~.$$
Therefore, the magnetic moment of $\Lambda$ - hyperon is equal to
$$\mu_\Lambda = \frac12(\mu_u - \mu_d + \mu_s - \mu_u + \mu_d +
 \mu_s) = \mu_s~~.$$
The separate consideration is also needed  for the transition
matrix element $\mu_{\Sigma \Lambda}$ that determines the
radiative decay $\Sigma^0 \rightarrow \Lambda +
\gamma$ (see the problem in the end of this lecture). In close analogy
to the proton wave function the wave function of $\Sigma^0$ - hyperon
is equal to
$$
{\dot \Sigma}^0 = \sqrt{\frac23} {\dot u}{\dot d}\d{s}-
\sqrt{\frac13}\sqrt{\frac12}({\dot u}\d{d} + \d{u}{\dot d}){\dot
s}~~.  $$
Then $\mu_{\Sigma \Lambda}$ is easily calculated
$$\mu_{\Sigma \Lambda} =  -\sqrt{\frac13}\frac12(\mu_u - \mu_d + \mu_s
 + \mu_u - \mu_d - \mu_s) = -\sqrt{\frac13}(\mu_u - \mu_d)~~.$$

The quark magnetic moments ($\mu_u, \mu_d, \mu_s$)
can be considered as parameters and defined, say, by magnetic moments
of proton, neutron and $\Lambda$ - hyperon
$$ \mu_u=\frac{4\mu_p+\mu_n}{5} = 1.85~~~~~~
\mu_d=\frac{4\mu_n+\mu_p}{5} = -0.97~~~~~~\mu_s=\mu_\Lambda=-0.61~~.$$
If $SU(3)$ - symmetry was exact symmetry the  quark magnetic moments
would be proportional to their electric charges
$$\mu_u : \mu_d : \mu_s = \frac23 : -\frac13 : -\frac13 = 2 : -1 :
-1~~.$$
As it can be seen the ratio $\mu_u : \mu_d = 1.85 : -0.97 \approx 2 : -1$
is satisfied rather well i.e. the $SU(2)$ - symmetry is a good symmetry
while the ratio $ \mu_d : \mu_s = -0.97 : -0.61 \neq -1 : -1$ 
is largely violated.\\
If $\mu_u : \mu_d = 2 : -1$ ($SU(2)$-symmetry) then $\mu_p : \mu_n = 3 : -2$
that is very close to the experimental result $\mu_p : \mu_n = 2.79 : -1.91$.
In such a way the additive quark model relates the 
magnetic moments of proton and neutron.

Finally, let us compare the predictions of $SU(3)$ - symmetry ,
the additive quark model and the experimental data
\vspace{1cm}

\begin{tabular}{cccc}
$\mu_p$~~~~~~~~~&input~~~~~~~~~ &input~~~~~~~~~&2.79\\
$\mu_n$~~~~~~~~~&input~~~~~~~~~ &input~~~~~~~~~&-1.91\\
$\mu_{\Sigma^+}$~~~~~~~~~&2.79~~~~~~~~~ & 2.67~~~~~~~~~&2.42\\
$\mu_{\Sigma^-}$~~~~~~~~~&-0.88~~~~~~~~~ & -1.09~~~~~~~~~&-1.16\\
$\mu_{\Xi^-}$~~~~~~~~~&-0.88~~~~~~~~~ & -0.49~~~~~~~~~&-0.68\\
$\mu_{\Xi^0}$~~~~~~~~~&-1.91~~~~~~~~~ & -1.43 ~~~~~~~~~&-1.25\\
$\mu_{\Lambda}$~~~~~~~~~&-0.96~~~~~~~~~ &input ~~~~~~~~~&-0.61\\
$\mu_{\Sigma\Lambda}$~~~~~~~~~&1.65~~~~~~~~~ &1.63~~~~~~~~~&1.61
\end{tabular}
\vspace{1cm}
 
It can be seen from the table that the additive quark model
gives better description of experimental data. Note that the
hypothesis of additivity is a natural but not an exact one.
Gluon corrections of quark interactions gives, for example, 
non additive contributions. These contributions make
the  description of experimental data better.

\subsection{Problem 8}

Calculate the width of $\Sigma^0 \rightarrow \Lambda + \gamma$
decay and find $\mu_{\Sigma \Lambda}$ from the comparison with
the experiment.

\newpage
\section{Lecture 9. Chiral symmetry and quark masses}

This lecture closes the round of lectures on the quark structure
of hadrons constructed from light quarks. We began with the introduction
of the masses of light $u,d,s$ - quarks ($4, 7, 150 MeV$) and said
that the difference of quark masses is small compared to the
characteristic hadronic scale $1 GeV$. From this followed the isotopic
($SU(2)$) and unitary ($SU(3)$) symmetries of strong interaction.
We considered the violation of $SU(3)$ - symmetry caused by the
difference of strange and nonstrange quark masses. Now it is quite
a time to say where are the quark masses used overall the lectures
came from. For this we will consider the so called chiral symmetry
of strong interaction corresponding to the case of massless quarks.
We have all the reasons to expect that the chiral symmetry is violated
no stronger than $SU(3)$ - symmetry based on the neglection of
strange and nonstrange quark mass difference.  Analyzing the
violation of the chiral symmetry we will find the quark masses.

\subsection{Left and right quarks}

If to put the quark masses to zero two nontrivial things happen:
the Lagrangian of strong interaction becomes invariant under
the separate $SU(3)$ - transformations of left and right (in the sense
of chirality) quark fields; left and right (in the sense
of chirality) quark fields goes in one to one correspondence
with  left and right (in the sense of spirality) states
of massless quarks. The symmetry for which left and right
quark fields transform separately is called the chiral symmetry.
In our case it is the chiral $SU(3)_L\times SU(3)_R$ - symmetry.
This symmetry means that  the left and right quark currents 
are conserved separately 
$$j^a_{\mu L} = {\bar q}_{L}\gamma_\mu \frac{\lambda^a}{2} q_L~~~~~~
j^a_{\mu R} = {\bar q}_{R}\gamma_\mu \frac{\lambda^a}{2} q_R,$$
where $q_{L,R} = \frac12 (1\pm\gamma_5)q$. The conservation of
quark current that is the sum of the left and right quark currents 
(vector current) corresponds to the  $SU(3)$ - symmetry considered
in lecture 5. 
The conservation of
quark current that is the difference of the left and right quark currents 
(axial current) corresponds to the  axial $SU(3)$ - symmetry. These
currents are
$$j^a_{\mu V} = {\bar q}\gamma_\mu
\frac{\lambda^a}{2} q~~~~~~ j^a_{\mu A} = {\bar q}\gamma_\mu\gamma_5
\frac{\lambda^a}{2} q.$$
Vector and axial currents generates conserved charges $Q^a, Q^a_A$.
The charges $Q^a$ are the charges that are  conserved due to
the $SU(3)$ - symmetry. In the limit of massless quarks the
symmetry is enlarged and new conserved charges - axial charges $Q^a_A$
have appeared.

\subsection{Nonlinear realization of chiral symmetry}

Let us consider the matrix elements of conserved vector and axial
currents.\\
Let the vector current be, for example, an  electromagnetic one
$$j^{em}_\mu = \frac23{\bar u}\gamma_\mu u - \frac13{\bar
d}\gamma_\mu d - \frac13{\bar s}\gamma_\mu s =
{\bar q}\gamma_\mu \frac{\lambda^{em}}{2} q,$$
where $\frac{\lambda^{em}}{2} = \frac{\lambda^{3}}{2} +
\frac{\lambda^{8}}{2\sqrt{3}}$, and matrix element be taken over
the proton states. In the limit of zero momentum transfer
$q \rightarrow 0$ the considered matrix element is equal to
$$<p(k_2)|j^{em}_\mu|p(k_1)> = {\bar \Psi}_2\gamma_\mu \Psi_1~~.$$
The conservation of  electromagnetic current means that
$$(k_{2\mu} - k_{1\mu}){\bar \Psi}_2\gamma_\mu \Psi_1 = 0~~.$$
This equation is valid due to the Dirac equations for proton
spinors $\Psi_2$ and $\Psi_1$.\\
Let now the axial current be the current corresponding to 
$\beta$ - decay of neutron
$$\gamma_\mu \rightarrow \gamma_\mu \gamma_5 ~~~~~~
\lambda^{em} \rightarrow \tau^+.$$
If in analogy to the case of electromagnetic current we assume that the
matrix element of the axial current over the states of neutron and
proton and in the limit of zero momentum transfer is equal to
$$<p(k_2)|j^{+}_{\mu
A}|n(k_1)> = g_A{\bar \Psi}_2\gamma_\mu \gamma_5\Psi_1~~,$$
then the conservation of the axial current will require the equation
$$(k_{2\mu} - k_{1\mu}){\bar \Psi}_2\gamma_\mu
\gamma_5\Psi_1 = g_A2m_p{\bar \Psi}_2\gamma_5\Psi_1 = 0~~$$
and hence will require the mass of nucleon be equal to zero $m_p = 0$.
With experimental mass of nucleon of the order $1 GeV$ we will be
forced to conclude that the chiral symmetry is strongly violated.\\
We will assume that the 
matrix element of the axial current over the states of neutron and
proton and in the limit of zero momentum transfer is equal to
$$<p(k_2)|j^{+}_{\mu A}|n(k_1)> = g_A{\bar
\Psi}_2\gamma_\nu \gamma_5\Psi_1(\delta_{\mu\nu}-\frac{q_\mu
q_\nu}{q^2})~~,$$
i.e. is conserved automatically. For $q^2 = 0$ the matrix element
has a pole that can be interpreted as follows: neutron emits
massless pseudoscalar particle ($\pi^-$) and transforms to proton
but $\pi^-$ is annihilated by axial current.\\
Eight massless pseudoscalar mesons correspond to eight conserved
axial currents. The situation when the masses of pseudoscalar mesons
are equal to zero in the chiral limit of massless quarks seems
to be more interesting from the physical point of view because
experimentally all eight pseudoscalar mesons are substantially
lighter than other hadrons. Especially it is true for the case of
$\pi$ - mesons. So, the chiral symmetry can be realized by
making massless only pseudoscalar mesons.\\
Vector charges $Q^a$ acting on the states of baryon octet
transform these states into themselves. Axial charge $Q^3_A$, for
example, acting, say, on the proton state transforms the proton state
to another state that has the same spin, isospin, hypercharge but
has opposite parity. In the octet of baryons there is no such states,
it is natural to think that we get the two particle state
"$\pi +$ nucleon". For zero energy of $\pi$ - meson the states
"nucleon" and "$\pi +$ nucleon" are degenerate as it should be
in the case when there is a symmetry. The described realization
of chiral symmetry is called the nonlinear realization.

\subsection{Partial conservation of axial currents and
quark masses}

The arguments in favor of zero masses of pseudoscalar mesons
for the limit of massless r can be obtained also by the
considerations of matrix elements of axial currents over
the pseudoscalar meson states and vacuum. For example, for $\pi^+$ -
meson the matrix element
$$<0| {\bar
 u}\gamma_\mu\gamma_5 d|\pi^+> = f_{\pi}p^{\pi^+}_\mu$$ 
enters the amplitude of weak decay $\pi^+\rightarrow \mu^+ { \nu}_\mu$.
Calculating the divergence of axial current we obtain
 $$<0| (m_u+m_d){\bar u}\gamma_5 d|\pi^+> =
 f_{\pi}m^2_\pi,$$
from where it can be seen that the mass of $\pi$ - meson goes to
zero in the limit of massless quarks $m_{u,d}\rightarrow 0.$
Similarly
 $$<0| (m_u+m_s){\bar u}\gamma_5 s|K^+> =
 f_{K}m^2_K~~.$$
Assuming that $<0| {\bar u}\gamma_5 d|\pi^+> \approx
<0| {\bar u}\gamma_5 s|K^+> $ and  $f_\pi \approx f_K$
($f_K = 1.25
f_\pi$ experimentally) we get the following quark mass ratio
$$\frac{m_u + m_d}{m_u + m_s} \approx \frac{m^2_\pi}{m^2_K}\approx
\frac{1}{13}~~.$$
If one takes the strange quark mass to be $m_s \approx 150 MeV$
(the mass differences of baryons in the decuplet  
that differ by the numbers of strange quarks are approximately
of this value) then
$$m_u + m_d \approx 11 MeV~~.$$
Now we have to evaluate the quark masses $m_u$ and $m_d$ 
separately. Up to now we did not take into account the
electromagnetic interaction that can give the electromagnetic
corrections to the masses of mesons. The size of these corrections
is comparable to the masses of $u$ and $d$ - quarks. Therefore,
our wishes to evaluate the quark masses $m_u$ and $m_d$ 
separately encounter the necessity to consider the
 electromagnetic corrections. There is no rigorous way to
 do it and we will bound ourselves by the following speculation.
In the spirit of above relations among the masses of mesons
squared and quark masses let us write the modified relations
that will take into account the difference of charged particles
from  ones
$$\begin{array}{l}
m^2_{\pi^+} \sim m_u + m_d + \gamma\\
m^2_{\pi^0} \sim m_u + m_d \\
m^2_{K^+} \sim m_u + m_s + \gamma\\
m^2_{K^0} \sim m_d + m_s
\end{array} ~.      $$
The unknown electromagnetic correction $\gamma$ can be eliminated
and we get as a result
$$\frac{m_d - m_u}{m_d + m_u} = \frac
{(m^2_{\pi^+}-m^2_{\pi^0}) - (m^2_{K^+}-m^2_{K^0})}{m^2_{\pi^0}}
\approx 0.29~~.$$
So, for $m_u + m_d \approx 11 MeV$ one obtains
$m_u \approx 4 MeV$ and $m_d \approx 7 MeV$. \\
We see that $m_d -m_u \sim m_d \sim m_u$, hence, the chiral
$SU(2)_L\times SU(2)_R$ - symmetry should be as good as
usual isotopic $SU(2)$ - symmetry and be better than
$SU(3)$ - symmetry in any case.

\subsection{Problem 9}

Find the relation between the ratio $\frac{m^2_{\pi^0}}{m^2_{\eta_8}} $
and the ratio of quark masses and verify its accuracy for the
values of quark masses obtained on the lecture.

\end{document}